\documentclass[preprint]{ptephy_v1}
\preprintnumber{XXXX-XXXX}

\usepackage{graphicx}
\usepackage[utf8]{inputenc}
\usepackage[separate-uncertainty]{siunitx}
\usepackage{amsmath}
\usepackage{color}
\usepackage{ulem}
\usepackage{subcaption}
\usepackage{multirow}

\begin{document}

\title{High-pressure xenon gas time projection chamber with scalable design and its performance at around the Q value of $^{136}$Xe double-beta decay}

\author[1,*]{Masashi~Yoshida}
\author[1]{Kazuhiro~Nakamura}
\author[2]{Shinichi~Akiyama}
\author[3]{Sei~Ban}
\author[1]{Junya~Hikida}
\author[1]{Masanori~Hirose}
\author[2]{Atsuko~K.~Ichikawa}
\author[4]{Yoshihisa~Iwashita}
\author[1]{Yukimasa~Kashino}
\author[1]{Tatsuya~Kikawa}
\author[5]{Akihiro~Minamino}
\author[6]{Kentaro~Miuchi}
\author[8]{Yasuhiro~Nakajima}
\author[2]{Kiseki~D.~Nakamura}
\author[1]{Tsuyoshi~Nakaya}
\author[9]{Shuhei~Obara\thanks{Present Address: Institute for Advanced Synchrotron Light Source, National Institute for Quantum Science and Technology, Sendai 980-8579, Japan}}
\author[7]{Ken~Sakashita}
\author[10,11]{Hiroyuki~Sekiya}
\author[2]{Hibiki~Shinagawa}
\author[1]{Bungo~Sugashima}
\author[2]{Soki~Urano}

\affil[1]{Department of Physics, Graduate School of Science, Kyoto University, Kyoto 606-8502, Japan \email{yoshida.masashi.8m@kyoto-u.ac.jp}}
\affil[2]{Department of Physics, Graduate School of Science, Tohoku University, Sendai 980-8578, Japan}
\affil[3]{International Center for Elementary Particle Physics, University of Tokyo, Tokyo, 113-0033, Japan}
\affil[4]{Institute for Integrated Radiation and Nuclear Science, Kyoto University, Kumatori 590-0494, Japan}
\affil[5]{Faculty of Engineering, Yokohama National University, Yokohama 240-8501, Japan}
\affil[6]{Department of Physics, Graduate School of Science, Kobe University, Kobe 657-0013, Japan}
\affil[7]{High Energy Accelerator Research Organization (KEK), Tsukuba 305-0801, Japan}
\affil[8]{Department of Physics, Graduate School of Science, University of Tokyo, Tokyo, 113-0033, Japan}
\affil[9]{Research Center for Neutrino Science, Frontier Research Institute for Interdisciplinary Sciences, Tohoku University, Sendai, 980-8578, Japan}
\affil[10]{Kamioka Observatory, Institute for Cosmic Ray Research, The University of Tokyo, Hida, 506-1205, Japan}
\affil[11]{Kavli Institute for the Physics and Mathematics of the Universe, The University of Tokyo, Kashiwa, 277-8583, Japan}

\begin{abstract}
  We have been developing a high-pressure xenon gas time projection chamber (TPC) to search for neutrinoless double beta ($0\nu\beta\beta$) decay of $^{136}\mathrm{Xe}$.
  The unique feature of this TPC is in the detection part of ionization electrons, called ELCC.
  ELCC is composed of multiple units, and one unit covers \SI{48.5}{\cm^2}.
  A \SI{180}{\L} size prototype detector with 12 units, 672 channels, of ELCC was constructed and operated with \SI{7.6}{bar} natural xenon gas to evaluate the performance of the detector at around the Q value of $^{136}\mathrm{Xe}$ $0\nu\beta\beta$.
  The obtained FWHM energy resolution is \SI{0.73+-0.11}{\%} at \SI{1836}{\keV}.
  This corresponds to \SI{0.60+-0.03}{\%} to \SI{0.70+-0.21}{\%} of energy resolution at the Q value of $^{136}\mathrm{Xe}$ $0\nu\beta\beta$.
  This result shows the scalability of the AXEL detector with ELCC while maintaining high energy resolution.
  Factors determining the energy resolution were quantitatively evaluated and the result indicates further improvement is feasible.
  Reconstructed track images show distinctive structures at the endpoint of electron tracks, which will be an important feature to distinguish $0\nu\beta\beta$ signals from gamma-ray backgrounds.
\end{abstract}
\subjectindex{H20}
\maketitle

\section{Introduction}
  Whether neutrinos have the Majorana nature is key to resolving the problems of the light neutrino masses\cite{MINKOWSKI1977421, yanagida1979, gellman1979} and the matter-antimatter asymmetry of the universe\cite{Fukugita:1986hr}.
  The most practical way considered so far to confirm the Majorana nature of neutrinos is to search for neutrinoless double-beta ($0\nu\beta\beta$) decay\cite{JJGOMEZCADENAS2012, doi:10.1146/annurev-nucl-101918-023407}.
  The current most stringent limit was obtained with the $^{136}\mathrm{Xe}$ nucleus; the KamLAND-Zen experiment set the lower limit of the half-life to be \num{2.3e26}~years (90\% C.L.)\cite{PhysRevLett.130.051801}.

  More sensitive searches for the $0\nu\beta\beta$ require a large target mass over ton-scale, ultra-low radioactivity in surrounding materials, and powerful discrimination between signals and backgrounds.
  In the case of $^{136}\mathrm{Xe}$, $2\nu$-emitting double-beta decay ($2\nu\beta\beta$) and gamma rays from $^{214}\mathrm{Bi}$ (\SI{2448}{\keV}) and $^{208}\mathrm{Tl}$ (\SI{2615}{\keV}) would be severe sources of backgrounds because they have close energy to the Q value of $0\nu\beta\beta$ (\SI{2458}{\keV}).
  High energy resolution is therefore essential for signal-background discrimination.
  Detection of the event pattern is also important because $0\nu\beta\beta$ has one cluster with two thick endpoints corresponding to two beta rays, whereas a gamma-ray event has multiple clusters or only one thick endpoint.
  A high-pressure xenon gas time projection chamber (TPC) has the potential to achieve these requirements.
  The application of high-pressure xenon gas TPCs for $0\nu\beta\beta$ searches began with the Gotthard experiment\cite{IQBAL1987459, LUESCHER1998407}.
  Now, there is leading research by the NEXT experiment\cite{NEXT:2020amj}, and the PandaX-III experiment\cite{Lin_2018} is also pursuing studies.

  We have been also developing a high-pressure xenon gas TPC named AXEL (A Xenon ElectroLuminescence detector) to search for $0\nu\beta\beta$.
  A peculiar feature of the AXEL detector is a unique counting technique of ionization electrons using electroluminescence (EL), called Electroluminescence Light Collection Cell (ELCC).
  We demonstrated the proof-of-principle of ELCC in Ref.\cite{BAN2017185} and showed the energy resolution of $1.73\pm0.07\%$~(FWHM) for \SI{511}{\keV} electrons in Ref.\cite{ban2020}.
  In this paper, we describe the performance at around the Q value of $^{136}\mathrm{Xe}$ $0\nu\beta\beta$, \SI{2458}{\keV}, with a larger-scaled AXEL prototype detector.

\section{Detector}
\begin{figure}[tb]
    \centering
    \includegraphics[width=0.95\linewidth]{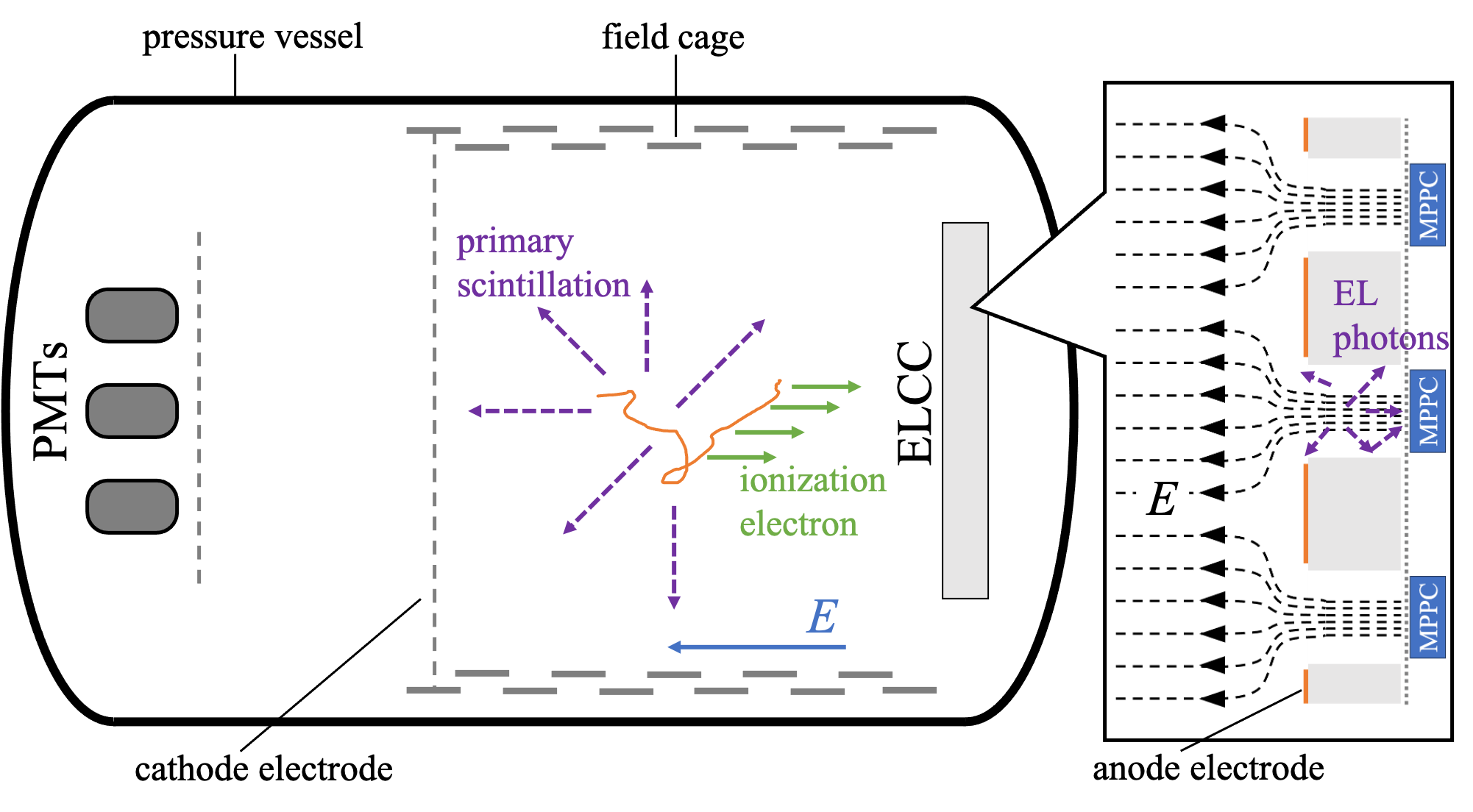}
    \caption{Schematic view of the AXEL \SI{180}{\L} prototype detector.}
    \label{fig:detector_overview}
  \end{figure}

  A schematic view of the AXEL detector is shown in Fig.~\ref{fig:detector_overview}.
  High energy charged particles deposit their energies by exciting and ionizing xenon atoms along with the tracks.
  Excited atoms emit the primary scintillation lights in hundreds of nanosecond time scales, and they are detected by photomultiplier tubes (PMTs) at the cathode side.
  Ionization electrons drift under the uniform electric field made by the field cage and are converted to photons and detected by the ELCC at the anode side in tens of microsecond time scales.
  The energies deposited by the charged particles are reconstructed from the photon counts at ELCC.
  The tracks of the charged particles are reconstructed from the hit pattern on the ELCC and the time difference between the hits of PMTs and ELCC.
  A large area ELCC is composed of multiple units.
    
  A 180 L prototype detector with three \SI{48.5}{\cm^2} ELCC units was constructed and evaluated in \cite{ban2020}.
  Since then, the sensitive volume of the detector was enlarged and the structure of ELCC was improved.

  \subsection{Electroluminescence light collection cell}
    ELCC is a pixelized detector for ionization electrons. It consists of the TPC-anode electrode, a ground potential mesh electrode, and a polytetrafluoroethylene plate (PTFE body) in between them.
    The anode electrode and PTFE body have holes ("cells") arranged in a hexagonal lattice pattern.
    An electric field up to \SI{3}{\kV/\cm/bar}, which is produced by the anode and mesh electrode and 30 times more intense than the drift field, draws ionization electrons into these cells and accelerates them.
    At this field, electrons excite but not ionize xenon atoms.
    EL photons are generated by the deexcitation of these atoms.
    For each cell, VUV-sensitive Silicon photomultipliers (Hamamatsu MPPC, S13370-3050CN) are placed behind the mesh electrode and detect the EL photons (See Fig.\ref{fig:detector_overview}).
    The advantages of ELCC are the following two points.
    One is that the number of detected photons is less dependent on the initial position of the ionization electrons because the EL process occurs after the ionization electrons are drawn into cells.
    The other is that there is little deformation of electrodes because the structure is supported by the PTFE body.

\begin{figure}[tb]
        \centering
        \includegraphics[width=0.8\linewidth]{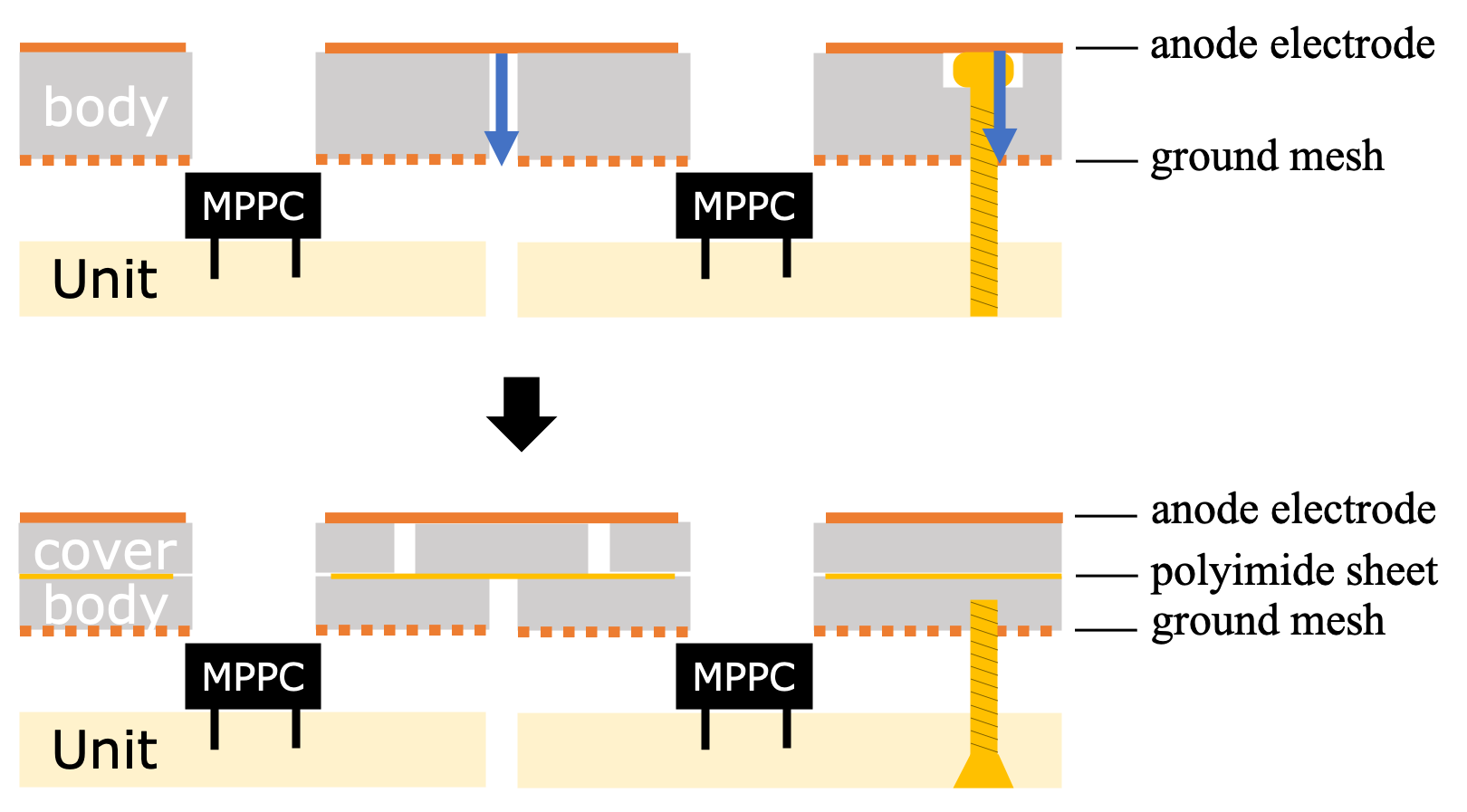}
        \caption{Schematic cross-sectional views of ELCC structures. The previous structure is shown in the upper part and the upgraded structure to prevent discharges is shown in the lower part. The paths of discharges are shown in the blue arrow in the upper part.}
        \label{fig:elcc_schematic}
    \end{figure}

    The plane of ELCC is made of parallelogram-shaped units of $56(=7\times8)$ channels each.
    The fundamental dimensions of ELCC are optimized\cite{ban2020} for the drift field of \SI{100}{\V/\cm/bar} and the EL field of \SI{3}{\kV/\cm/bar}, that is,
    \begin{itemize}
      \item Anode hole diameter: \SI{5.5}{\mm}
      \item PTFE body hole diameter: \SI{4.5}{\mm}
      \item Cell depth: \SI{5}{\mm}
      \item Cell pitch: \SI{10}{\mm}
    \end{itemize}
    The cross-sectional view of the ELCC plane with the previous setup adopted in \cite{ban2020} is shown in the upper part of Fig.~\ref{fig:elcc_schematic}.
    
\begin{figure}[tb]
      \centering
      \includegraphics[width=0.8\linewidth]{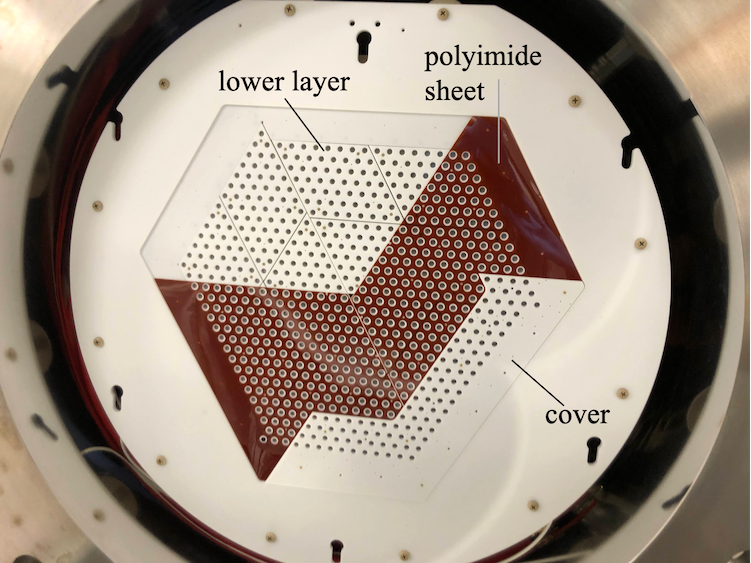}
      \caption{Photograph of the upgraded ELCC plane during assembly. Part of the polyimide sheets and the covers were removed.}
      \label{fig:elcc}
    \end{figure}

\begin{figure}[tb]
        \centering
        \includegraphics[width=0.8\linewidth]{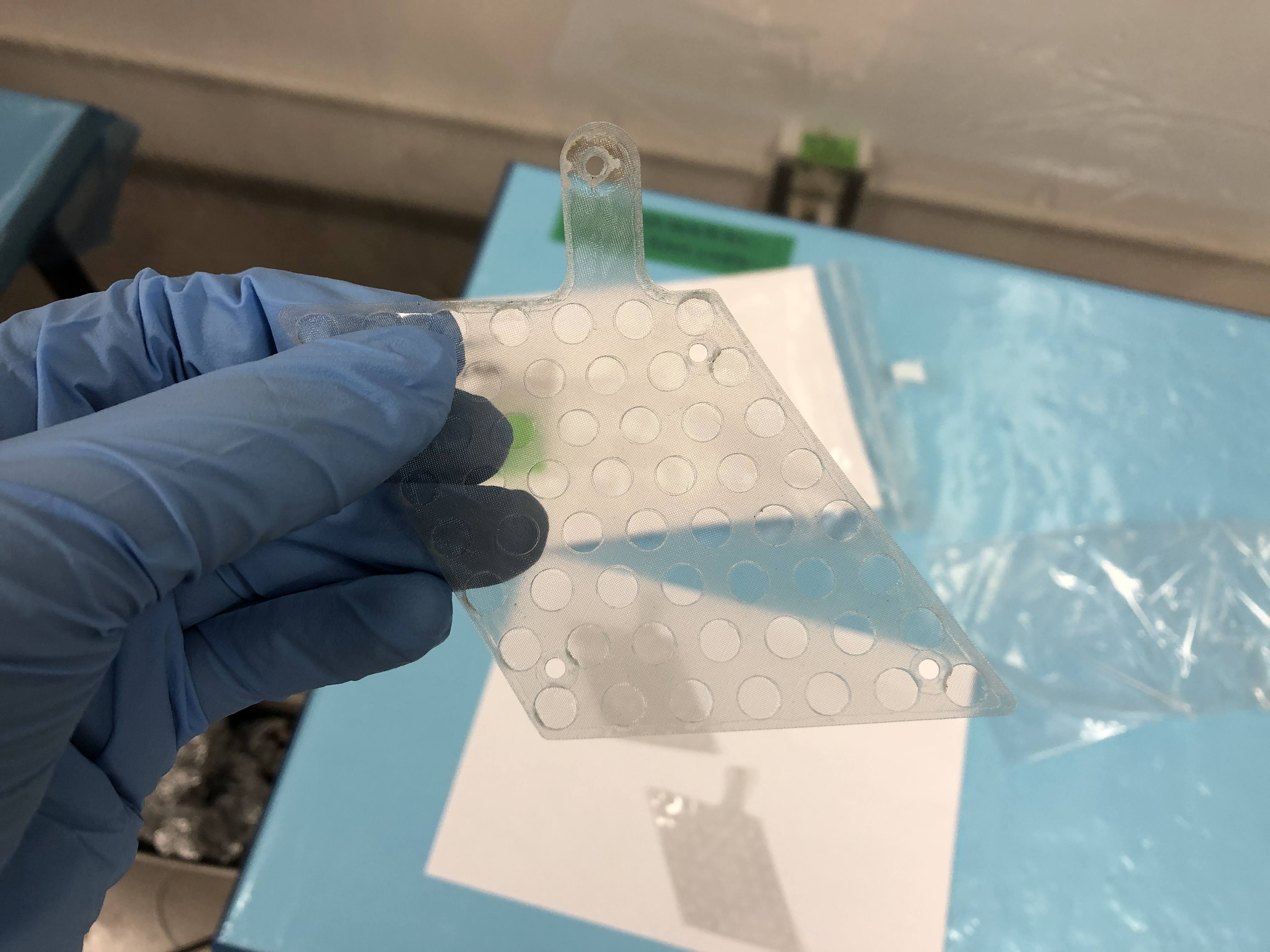}
        \caption{Photograph of the mesh electrode welded between PFA films.}
        \label{fig:mesh}
    \end{figure}

    There happened frequent and severe electric discharges when the operating voltage was increased together with the increase of operating pressure from \SI{4}{bar} to \SI{7.6}{bar}.
    The discharges occurred between the anode and the ground mesh electrodes at the boundaries of ELCC units and the screw holes to fix ELCC units.
    We have applied the following countermeasures to prevent these discharges.
    First, the PTFE body is separated into two layers.
    We call the upper (the anode electrode side) layer as "cover".
    The shape of the cover was changed from that of the units to shift the boundaries from those of the lower layers.
    Then, the anode electrode and the ground mesh electrodes do not face each other directly.
    A \SI{125}{\um} polyimide sheet is inserted between the two layers to block the intersection of the boundaries (See Fig. \ref{fig:elcc}).
    Second, the direction of the screws to fix ELCC units is reversed so that there are no holes going through the ELCC units.
    These two countermeasures are illustrated in the lower part of Fig. \ref{fig:elcc_schematic}.
    Lastly, the mesh electrodes are covered by two perfluoroalkoxyalkane (PFA) films so as not to expose sharp edges that trigger corona discharges (See Fig. \ref{fig:mesh}).
    This also prevents the mesh from fraying, and it suppresses discharges caused by mesh fragments getting into the cells.
    The PFA films have holes corresponding to the cells to prevent charge-up.
    
    Together with these countermeasures, the number of ELCC units was increased from 3 to 12, and the total number of channels is 672, to observe the events with higher energies and longer tracks. The sensitive area is around \SI{580}{\cm^2}.
    
    The signals of MPPCs are transferred via cables of flexible printed circuits (FPCs), and recorded with dedicated front-end boards AxFEB\cite{kznakamura2020} at \SI{5}{MS/s} for each unit.
    The bias voltages for the MPPCs are also supplied by the same FPC cables.
   
    \subsection{Field cage}
    An intense and uniform drift electric field is necessary to achieve high energy resolution and fine track image since it suppresses the fluctuation of recombination, attachment, and diffusion of ionization electrons during the drift.
    On the contrary, the collection efficiency of the ionization electrons into the ELCC cells decreases if the drift field is not sufficiently low compared to the EL field.
    We adopted \SI{100}{\V/\cm/bar} for the drift field as a design value with allowed deviations of $\pm$~5\%.
    The energy resolution is expected to get worse below \SI{100}{\V/\cm/bar} because of the recombination of ionization electrons\cite{BOLOTNIKOV1997360}.

\begin{figure}[tb]
      \centering
      \includegraphics[width=0.8\linewidth]{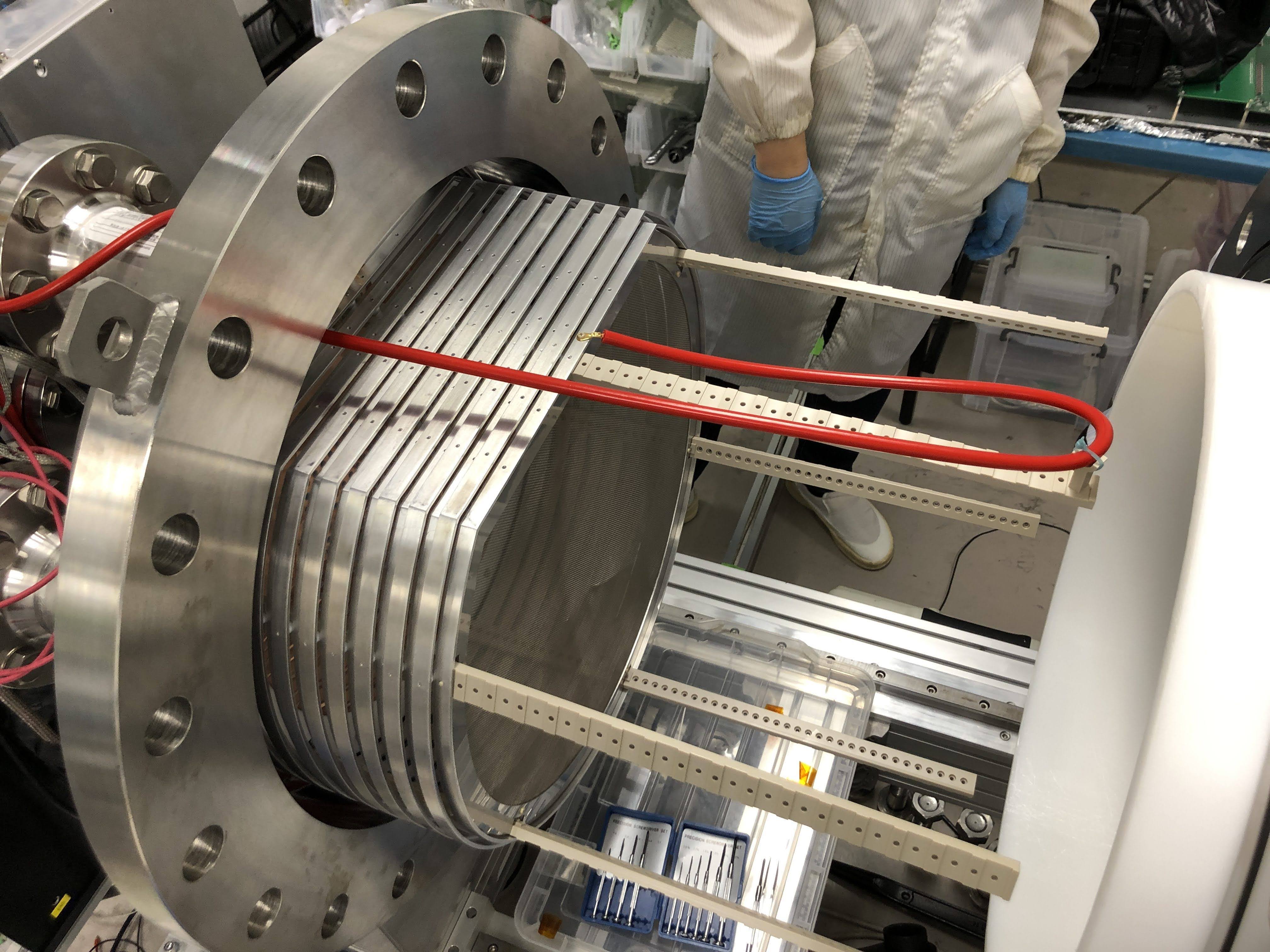}
      \caption{Photograph of the field cage installed in the pressure vessel. A white cylinder on the right side is the HDPE insulator.}
      \label{fig:fieldcage}
    \end{figure}

    The field cage to generate the drift field consists of \SI{3}{\mm} thick and \SI{12}{\mm} wide band-shaped aluminum electrodes aligned between the anode and the cathode electrodes.
    They have two different diameters (\SI{505}{\mm} and \SI{489}{\mm}) and are alternately lined up with an overlap of \SI{1}{mm} to shield the effect of the ground potential of the pressure vessel.
    The inner diameter of the pressure vessel is \SI{547}{\mm}.
    The vessel and the outer electrodes are insulated by a \SI{20}{\mm} thick high-density polyethylene (HDPE) cylinder.
    Each ring electrode has a straight section of \SI{300}{\mm} to make cabling space between the HDPE cylinder and the field cage.
    A cathode mesh electrode is placed on top of the ring electrode array.
    The mesh is point welded to a \SI{1}{\mm} thick stainless steel frame under tension, and thus deflection of the mesh is kept small.
    Negative high voltage is applied to the cathode mesh via a wire covered by silicone rubber.
    The cathode, ring electrodes, and anode are connected in series via \SI{100}{\Mohm} resistors to equally apply potential differences between neighboring electrodes.
    Six pillars made of polyetheretherketone (PEEK) support the electrodes from the inside.
    The ends of the pillars are fixed to the PTFE disk hosting the anode plane.
    Figure \ref{fig:fieldcage} is the photograph of the field cage.
    
    A finite element method calculation by FEMM\cite{femm} on this configuration showed that the intensity of the drift field satisfies \SI{100}{\V/\cm/bar}~$\pm$~5\% in the region within 229.3 mm from the central axis.
    The sensitive area of the 12-unit ELCC is fully inside this uniform region.

    The distance between the cathode mesh electrode and the anode plane, the drift length, is adjustable up to \SI{46}{\cm}.
    Due to the limitation of the cathode high voltage supply, however, it was limited to \SI{18}{\cm} in this paper.
    The sensitive volume is about \SI{10000}{\cm^3} as a result.

    \subsection{PMTs}
    We use VUV-sensitive and high-pressure tolerant PMTs, Hamamatsu R15298.
    The number of PMTs to detect primary scintillation lights is increased to 7 from 2 in the previous paper\cite{ban2020}.
    They are mounted behind the cathode mesh at a distance not to make an intense electric field over the EL threshold even if the field cage is fully extended to \SI{46}{\cm}.
    The resulting distance between PMTs and the cathode mesh this time is approximately \SI{38}{\cm}.
    A guard mesh at the ground potential is placed in front of the PMTs. The number of photons reaching the PMTs is decreased by the aperture ratio of the two meshes, 71\% for the cathode mesh and 67\% for the guard mesh.

    The signals of PMTs are transferred via PTFE-coated coaxial cables, amplified 100-fold with fast amplifiers, and recorded by a 100 MS/s waveform digitizer (CAEN, v1724).
    
\section{Measurement}\label{sec:measurement}
  To evaluate the performance of the upgraded detector at around the Q value of $^{136}\mathrm{Xe}$ $0\nu\beta\beta$, we conducted measurements with gamma-ray sources.
  The measurement conditions and procedure are described below.

  Before introducing xenon gas into the pressure vessel, an evacuation was conducted for two weeks.
  The vacuum level reached \SI{3.9e-2}{\Pa}, and the outgassing rate was \SI{1.23e-4}{\Pa.\m^3/\s}.
  After the evacuation, \SI{7.6}{bar} of natural xenon gas was filled, and the gas was circulated with a flow of \SI{5}{NL/min} and purified by a molecular sieve (Applied Energy Systems, 250C-V04-I-FP) and a getter (API, API-GETTER-I-RE).
  Before starting the measurement, we took three weeks of purification term while monitoring the improvement of the EL-light yield.

  The intensity of the EL and drift electric fields were \SI{2.5}{\kV/\cm/bar} and \SI{83.3}{\V/\cm/bar}, respectively.
  These are lower than the design values of \SI{3}{\kV/\cm/bar} and \SI{100}{\V/\cm/bar}.
  This is because frequent discharges still occurred at ELCC at the design value.
  The ratio between the EL and drift field intensity was kept to be 0.1:3 to maintain 100\% collection efficiency of ionization electrons into the ELCC cells.
  The applied high voltages were hence \SI{-9.5}{\kV} for the anode and \SI{-20.9}{kV} for the cathode.
  At these conditions, discharges took place once per several hours on both the anode and cathode.
  When a discharge occurs, an interlock system cuts off the high voltages, and they are re-applied manually.

  Two kinds of gamma-ray sources were used.
  One is an $^{88}\mathrm{Y}$ source, which mainly emits gamma rays of \SI{898.0}{\keV} and \SI{1836}{\keV}.
  The intensity of the source was \SI{9}{\kilo\becquerel}.
  Another one is a set of thoriated tungsten rods.
  They are commercial products for welding and include 2\% of thorium by mass.
  Thus they can be used as a source of thorium series radiations including \SI{2615}{\keV} gamma rays of $^{208}\mathrm{Tl}$.
  The amount of used thoriated tungsten rods was \SI{1}{\kg}, resulting in \SI{80}{\kilo\becquerel} of intensity.
  The source was set at the outside surface of the pressure vessel during the measurement.

\begin{figure}[tb]
    \centering
    \includegraphics[width=0.95\linewidth]{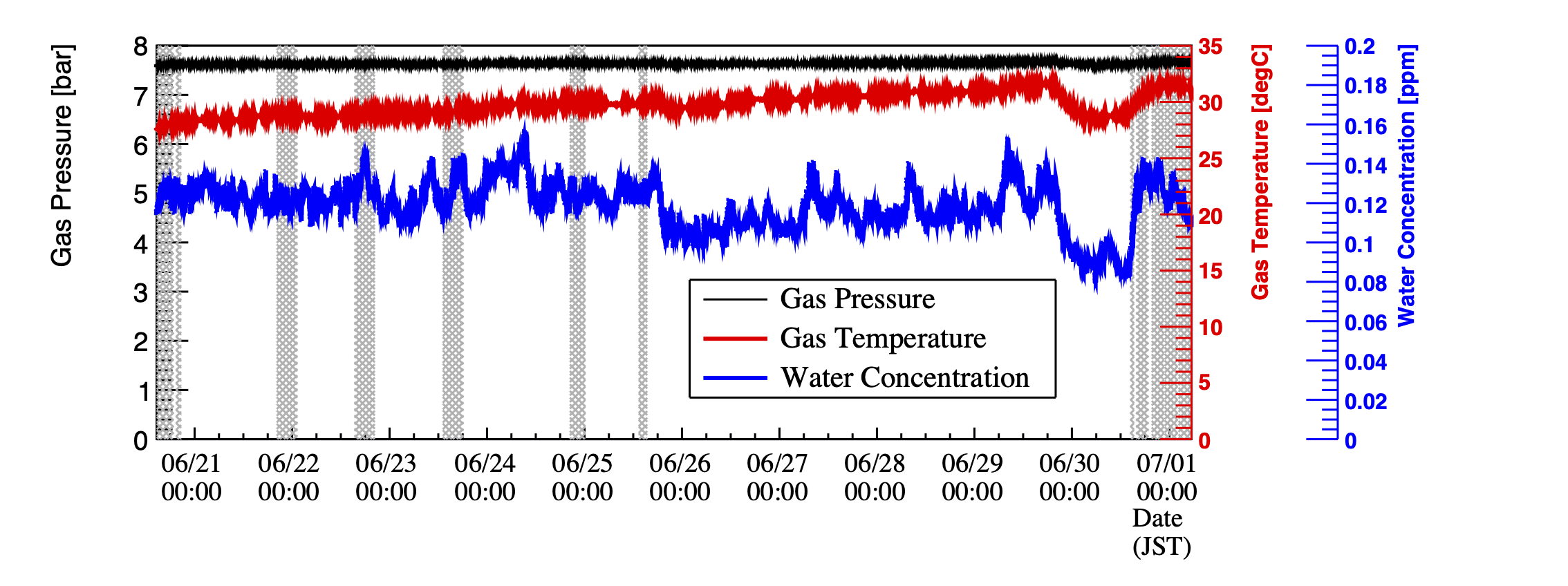}
    \includegraphics[width=0.95\linewidth]{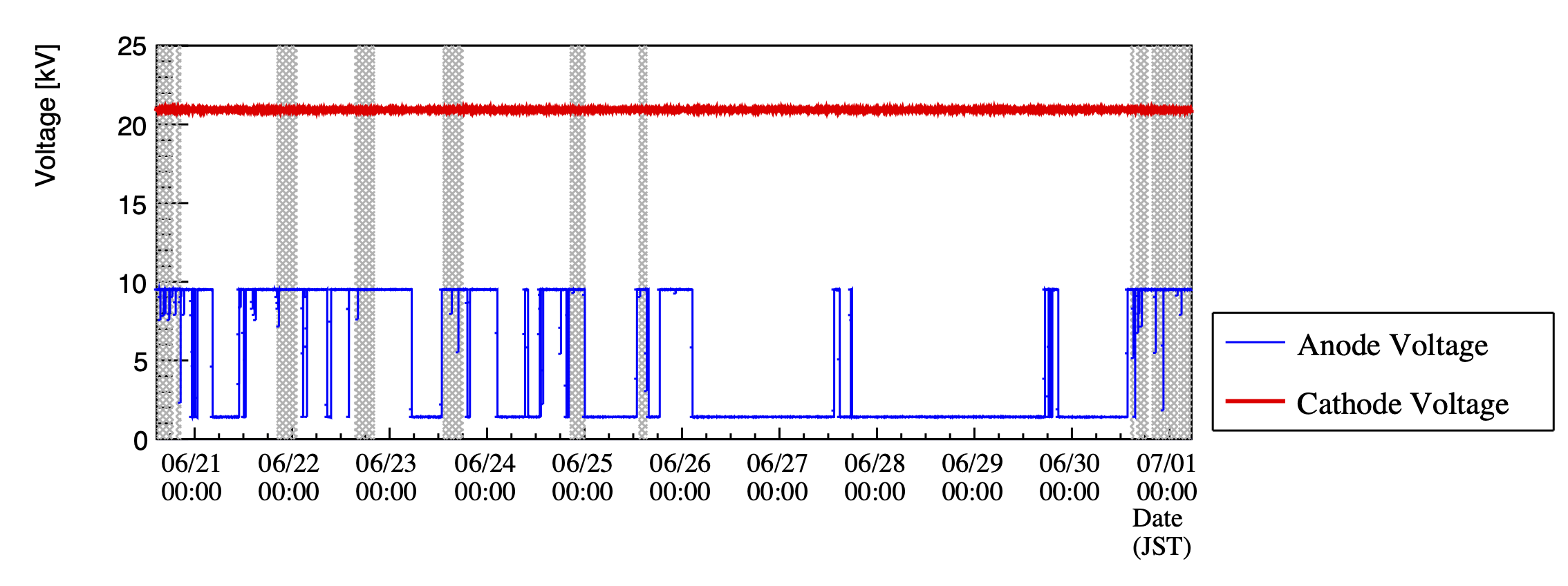}
    \caption{Trend of the monitor data. The upper panel is for gas conditions, and the lower panel is for high voltages. The gray-shaded areas are data-taking periods. The drop in the anode voltage corresponds to discharges or manual off.}
    \label{fig:monitor}
  \end{figure}

  Data were taken for five days for $^{88}\mathrm{Y}$ and one day for thoriated tungsten rods in June 2022 with intervals for commissioning and data checking.
  The anode and cathode voltages, gas pressure, gas temperature, water concentration, etc. were monitored during the data taking.
  Figure \ref{fig:monitor} shows the trends of important monitor values.

\begin{figure}[tb]
    \centering
    \includegraphics[width=0.8\linewidth]{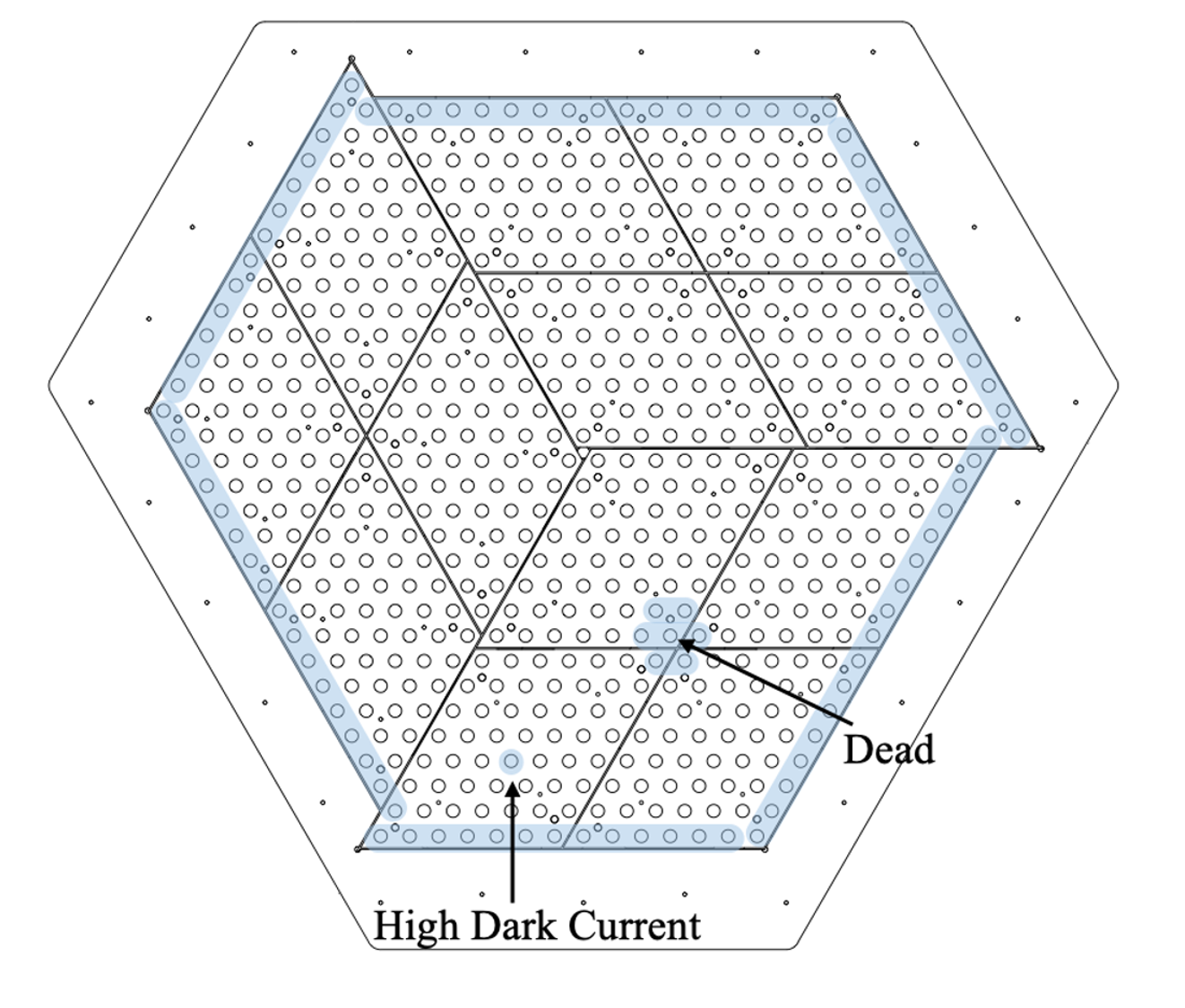}
    \caption{Configuration of veto channels. The blue-shaded channels were assigned to veto.}
    \label{fig:veto_channels}
  \end{figure}

  The outermost 87 channels out of 672 channels of ELCC were assigned to the veto.
  Apart from that, there was one channel with a high dark current and one dead channel.
  The high dark-current channel and six channels around the dead channel were also added to the veto (See Fig. \ref{fig:veto_channels}).
  
  Two kinds of triggers were used for the data acquisition of the ELCC signal.
  One is called fiducial trigger, which is issued when the sum height of the signal of channels other than the predefined veto channels exceeds a threshold, and the veto channels have no hits.
  The threshold of the fiducial trigger was set around \SI{400}{\keV}.
  The other, the whole trigger, sets a threshold on the signal sum for all channels including the veto channels.
  It is to take calibration data targeting \SI{30}{\keV} characteristic X-rays and was prescaled to 1/50.

  To issue the triggers, a trigger board, Hadron Universal Logic module (HUL)\cite{R.Honda2016}, is used.
  Each FEB sends the waveform to HUL, and HUL sends the trigger and veto signal and a common 160 MHz clock to FEBs.
  HUL outputs two other NIM signals.
  One is called send-trigger signal that is synchronized with the trigger to FEBs.
  The other is called send-header signal which is synchronized with the timing of data transmission to the DAQ PC after waveform acquisition at FEB is complete.

  The data acquisition of the PMT signal is triggered by the send-header signal from HUL.
  The waveform recording window is set to \SI{600}{\us}, and the pre-trigger region is 95\% of the window so that the timing of primary scintillation is certainly included in the window.
  The waveform digitizer for PMTs also records the send-trigger signal from HUL, and the signals were used to match the timing of the corresponding events of the ELCC and PMTs.

  The number of total acquired events is \num{1145761} for the $^{88}\mathrm{Y}$ run and \num{869422} for the thoriated tungsten rod run.
  We used the whole dataset to evaluate the detector performance.

\section{Analysis}
  The analysis process is composed of three steps.

  The first is the ELCC waveform analysis (Sec.\ref{sec:elcc_analysis}), which consists of the search for hits, clustering, correction to the non-linearity of MPPCs, and correction to the gain of the EL process in each ELCC cell.
  From this step, we can estimate the number of ionization electrons, or, the energy, and the track pattern.

  The second is the PMT waveform analysis (Sec.\ref{sec:pmt_analysis}), which consists of the search for PMT hits, identification of the primary scintillation, and matching to the ELCC events.
  From this step, we can determine the time of the event relative to the ELCC signal and hence absolute position of each ionization along the direction of the drift (hereafter, $z$-position).

  The last step is overall cuts and corrections (Sec.\ref{sec:cuts_corrections}).

  The same analysis method was used for both the $^{88}\mathrm{Y}$ run and the thoriated tungsten rod run, however, the correction coefficients are different for each run.

  \subsection{ELCC waveform analysis}\label{sec:elcc_analysis}
\begin{figure}[tb]
      \centering
      \includegraphics[width=0.8\linewidth]{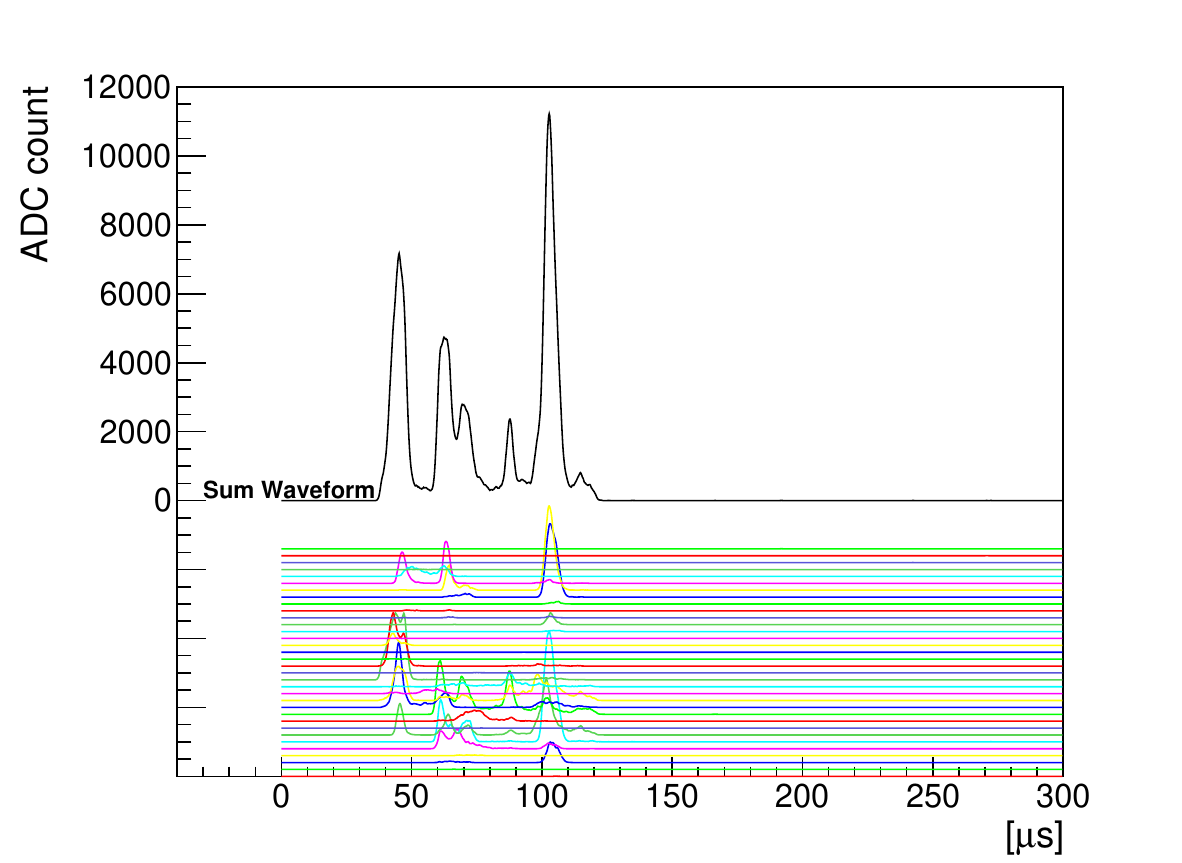}
      \caption{Typical waveforms of the ELCC signal. The waveform of each channel is drawn in colored lines with arbitrary offsets. The sum of the waveforms is drawn in the black line.}
      \label{fig:elcc_waveform}
    \end{figure}

    Figure \ref{fig:elcc_waveform} shows typical waveforms of the ELCC signal.
  
    Hits are searched for a threshold of \SI{3.5}{ADC} counts above the baseline, which corresponds to \SI{3.8}{photons} equivalent.
    The rise timing and the fall timing of a hit are defined as the threshold-crossing timing.
    From the five samplings before the rise timing to the five samplings after the fall timing, the ADC counts of the waveform are summed up, and the sum is converted to the photon count by using the gain of the channel's MPPC.
    The gains are pre-calculated using the MPPC's dark current pulses, as described in Refs.\cite{BAN2017185, kznakamura2020}.
    
    Hits that are in adjacent channels and overlapped in time are identified as belonging to the same single cluster.
    All hits in an event are assigned to clusters.

    Events are removed from further analysis if ADCs overflow, the rise timing is less than 20 samplings (\SI{4}{\us}) or the fall timing is over 1300 samplings (\SI{260}{\us}).
  
    \subsubsection{MPPC non-linearity correction}\label{sec:mppc_non_linearity}
      MPPCs have a non-linear output for high incident light intensity.
      This is because the number of MPPC pixels is limited, and it takes a finite time for each pixel to restore the bias voltage after the charge is released by photon detection.
      Thus, the non-linearity is characterized by the number of pixels and recovery time and is corrected with the following equation\cite{ban2020}:
      \begin{equation*}
        N_\mathrm{corrected} = \frac{N_\mathrm{observed}}{1-\frac{\tau}{\Delta t \cdot N_\mathrm{pixel}}N_\mathrm{observed}} ,
      \end{equation*}
      where $N_\mathrm{observed}$ and $N_\mathrm{corrected}$ are the photon counts before and after the correction respectively, $\tau$ is the recovery time, $\Delta t$ is the time width applying this correction, \SI{200}{\ns} in this analysis corresponding to the sampling speed of the FEBs, and $N_\mathrm{pixel} = 3600$, the number of pixels.

\begin{figure}[tb]
        \centering
        \includegraphics[width=0.8\linewidth]{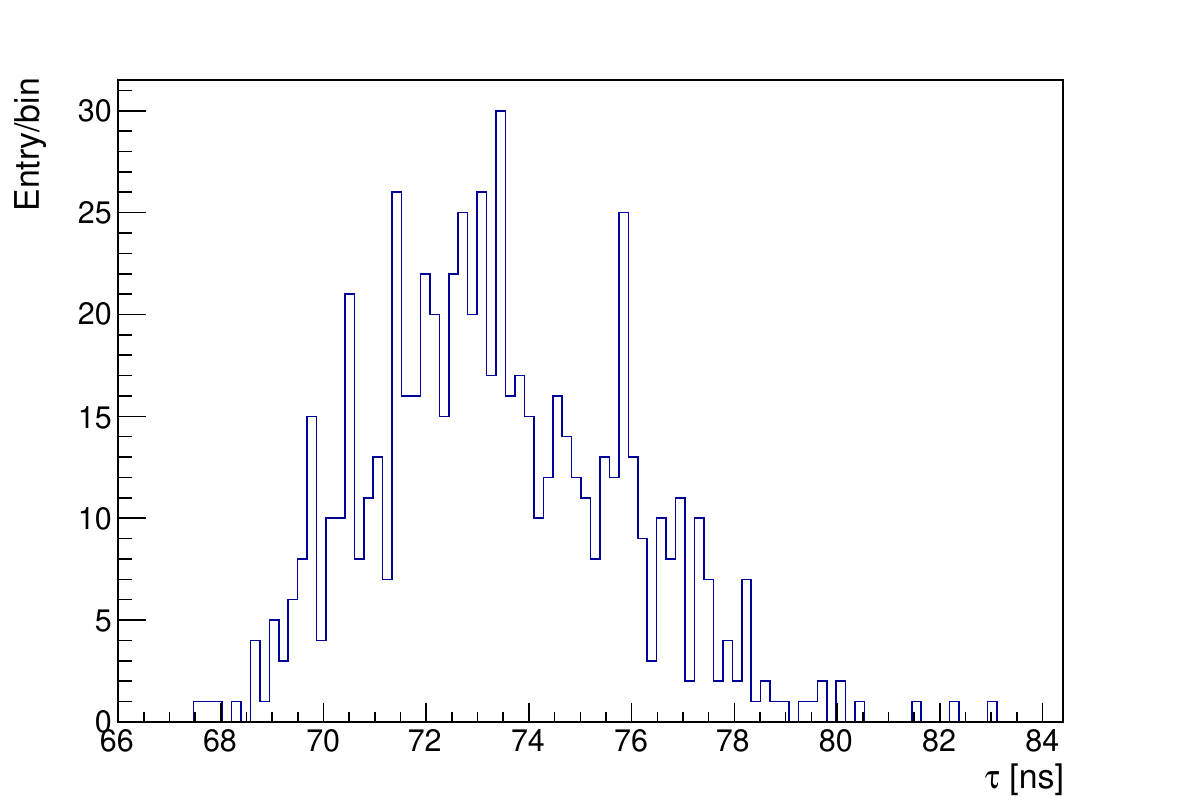}
        \caption{Distribution of the measured recovery times of MPPCs.}
        \label{fig:recovery_time_distribution}
      \end{figure}

      The recovery times of each MPPC were measured in advance by measuring the responses to the high-intensity LED light.
      To monitor the true number of photons incident to MPPCs, one MPPC with a 5\%-ND filter attached was used as a reference.
      The mean of the measured recovery times is \SI{73.4}{\ns}.
      Figure \ref{fig:recovery_time_distribution} shows the distribution of the measured recovery times.
    
    \subsubsection{EL gain correction}\label{sec:elgain}
      The gain of the EL process in each ELCC cell (hereafter, EL gain) is defined as the mean of detected photon count when one ionization electron enters the cell.
      The EL gains are different channel by channel due to dimensional differences by machining accuracy, differences in the photon detection efficiency of MPPCs, and so on.
      Thus, each signal should be corrected using its gain relative to the mean over channels to obtain better energy resolution.

      The correction factors are determined by using the peak of $\mathrm{K_\alpha}$ characteristic X-rays (\SI{29.63}{\keV}) in the photon count spectra.
      To conduct this, clusters in which the target channel has the highest photon counts are used.
      Throughout this procedure, the correction factors of the adjacent channels affect the determination of the correction factor of the channel.
      Hence this procedure is applied to all channels and iterated multiple times until the factors converge: in this analysis, six times.

      The mean of the EL gains is 12.5~photon/electron for this measurement.
    
  \subsection{PMT waveform analysis}\label{sec:pmt_analysis}
\begin{figure}[tb]
      \centering
      \includegraphics[width=0.95\linewidth]{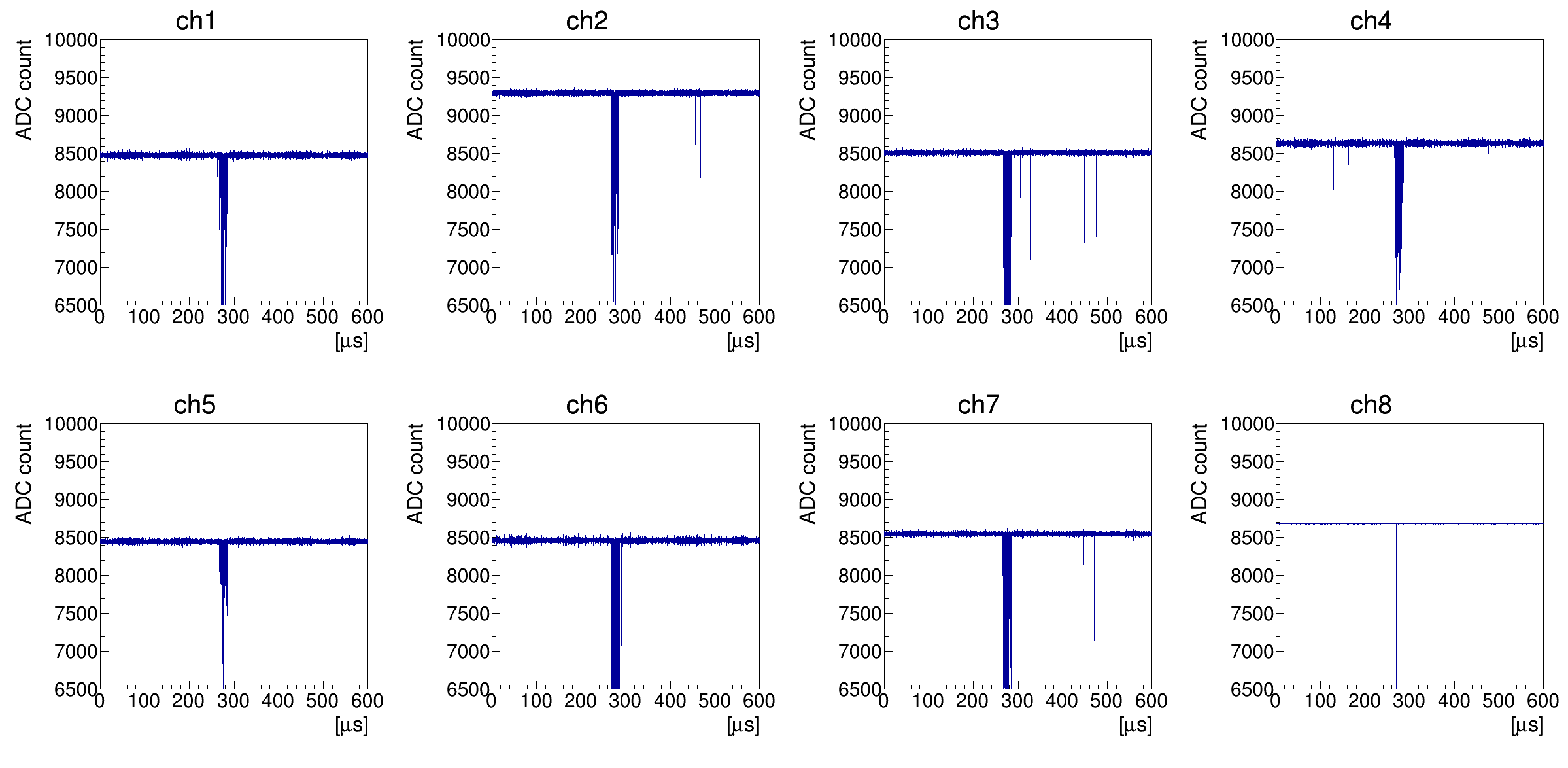}
      \caption{Typical waveforms of the PMTs. The hits at $\sim$\SI{130}{\us} in channel 4 and channel 5 are the scintillation signals. The hits at around \SI{300}{\us} are EL from ELCC. Channel 8 records the send-trigger signal from HUL as mentioned in Sec. \ref{sec:measurement}, not the PMT signal.}
      \label{fig:pmt_waveform}
    \end{figure}

    Figure \ref{fig:pmt_waveform} shows typical signal waveforms of the PMTs.
    Very narrow hits coinciding in two PMTs and preceding the EL signals are the primary scintillation photon candidates.

    The hit threshold is set at 200 ADC counts below the baseline.
    It is sufficiently higher than noise and lower than 1~p.e. wave height.
    To separate hits by scintillation light from hits by EL lights, hits are selected when they have a width less than \SI{400}{\ns} and are more than \SI{1}{\us} apart from other hits.
    Hereafter, hits selected by this criterion are called scintillation-like hits, and the others are called EL-like hits.
    Of these scintillation-like hits, those that are coincident within \SI{100}{\ns} in two or more channels are reconstructed as a hit cluster by the primary scintillation light.
    In case there are two or more hit clusters, such events are cut because it is not possible to determine which hit cluster is the actual primary scintillation.

\begin{figure}[tb]
      \centering
      \includegraphics[width=0.8\linewidth]{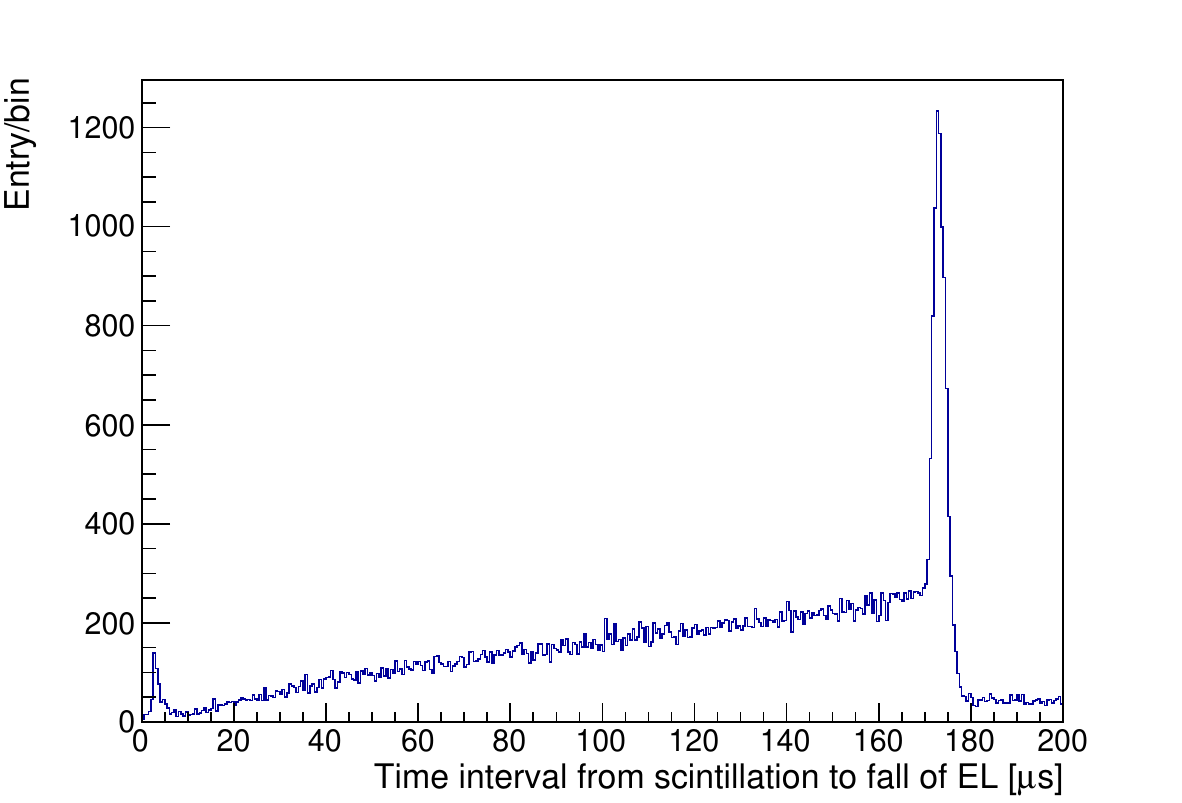}
      \caption{Distribution of the time intervals between scintillation and the fall timing of ELCC events. The peak at \SI{175}{\us} is formed by the events across the cathode plane.}
      \label{fig:fall_timing}
    \end{figure}

    The corresponding ELCC event and PMT event are matched based on the information of the timestamp and the internal clock of the ELCC FEBs and the PMT digitizer.
    For matched events, the time interval between the primary scintillation and the ELCC hits is calculated with the help of the send-trigger signal from HUL which corresponds to the fixed timing in the data acquisition window for ELCC.
    Figure \ref{fig:fall_timing} shows the distribution of the time intervals between the primary scintillation and the fall timing of ELCC events.
    The peak in this distribution corresponds to the cathode plane of the field cage: in other words, $z = \SI{18}{cm}$.
    From this, the drift velocity of ionization electrons was derived to be \SI{1.04}{\mm/\us}.
    Using this drift velocity, the $z$-position of each ionization signal detected by ELCC is reconstructed.

    \subsubsection{Selections to avoid timing mismatch}
      Events are cut when there are multiple send-trigger signals from HUL because it leads to a timing mismatch depending on which one corresponds to the true beginning of the ELCC events.
      This can be caused by the pile-up of events during the drift of ionization electrons.

      Events with EL-like hits within the \SI{18}{cm}-equivalent time preceding the ELCC signal are cut.

      Finally, events are cut if there exist scintillation-like hits, even a single hit without any coincidence, in the region corresponding to within \SI{2}{\cm} from the ELCC signal.
      This is because ionization electrons generated within \SI{2}{cm} from the ELCC surface are not guaranteed to have 100\% collection efficiency into the ELCC\cite{ban2020}, thus the energy resolution gets worse if it fails to cut events whose actual rising edge is within \SI{2}{\cm}.

  \subsection{Fiducial volume cuts and overall corrections}\label{sec:cuts_corrections}
    Using the information obtained from the ELCC and PMT signal analysis, fiducial volume cut and overall corrections are applied.
    
    \subsubsection{Elimination of clusters with small photon counts}
\begin{figure}[tb]
          \centering
          \includegraphics[width=0.8\linewidth]{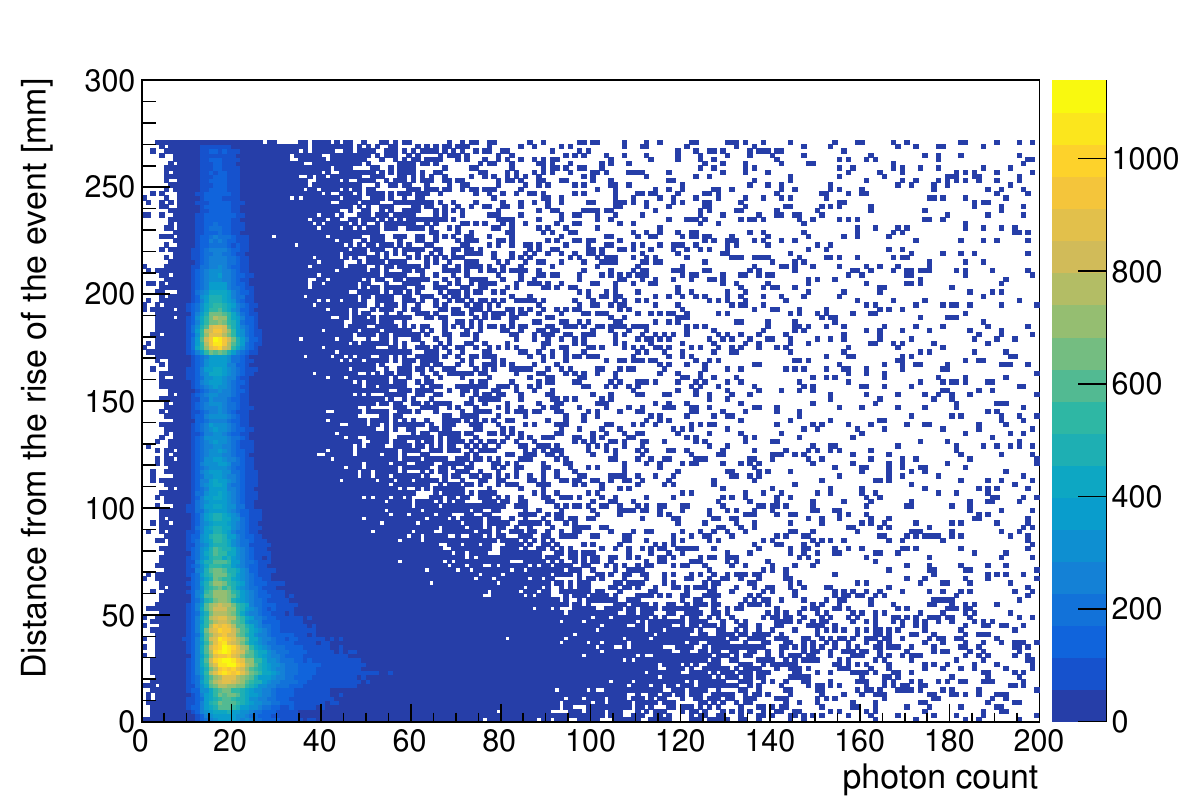}
          \caption{ Drift distance after the rising of the event and photon counts for the clusters with small photon counts.}
          \label{fig:small_photon_count_clusters}
      \end{figure}

      Many events contain a few to several tens of clusters with photon counts less than one hundred.
      These clusters are generated from one to a few electrons.
      Figure \ref{fig:small_photon_count_clusters} shows the distribution of the photon counts and the drift distance after the rising of the ELCC signal for these small photon count clusters.
      The dense region around \SI{180}{\mm}, which corresponds to the length of the drift region, is considered to be clusters by electrons generated by VUV EL-light hitting the cathode mesh.
      The other clusters can be formed in the same manner with various detector components and also by the ionization electrons that attach to impurities in the gas during drift and are released after a while.
      Such a phenomenon is also observed in liquid xenon detectors\cite{ZEPLIN-III:2011qer, Aprile_2014, Sorensen_2018}.

      Because these clusters with small photon counts disturb fiducial cut (Sec. \ref{sec:fiducial_cut}) and reconstruction of tracks, the clusters with less than 100 photons are eliminated from events.
      The total photon counts of eliminated clusters are 400 photons at maximum for each event.
      Thus the effect on the reconstruction of energies is less than 0.04\% for the \SI{1836}{\keV} photopeak and negligible compared to the energy resolution.
      
    \subsubsection{Fiducial volume cut}\label{sec:fiducial_cut}
      Fully contained events are selected by rejecting events that have any hits on veto channels (Sec.\ref{sec:measurement}) and events whose $z$-position extends beyond the $\SI{2}{\cm}<z<\SI{17.5}{\cm}$ region.
      
    \subsubsection{Correction of time variation}\label{sec:time_correction}
\begin{figure}[tb]
        \begin{minipage}{0.5\linewidth}
          \centering
          \includegraphics[width=\linewidth]{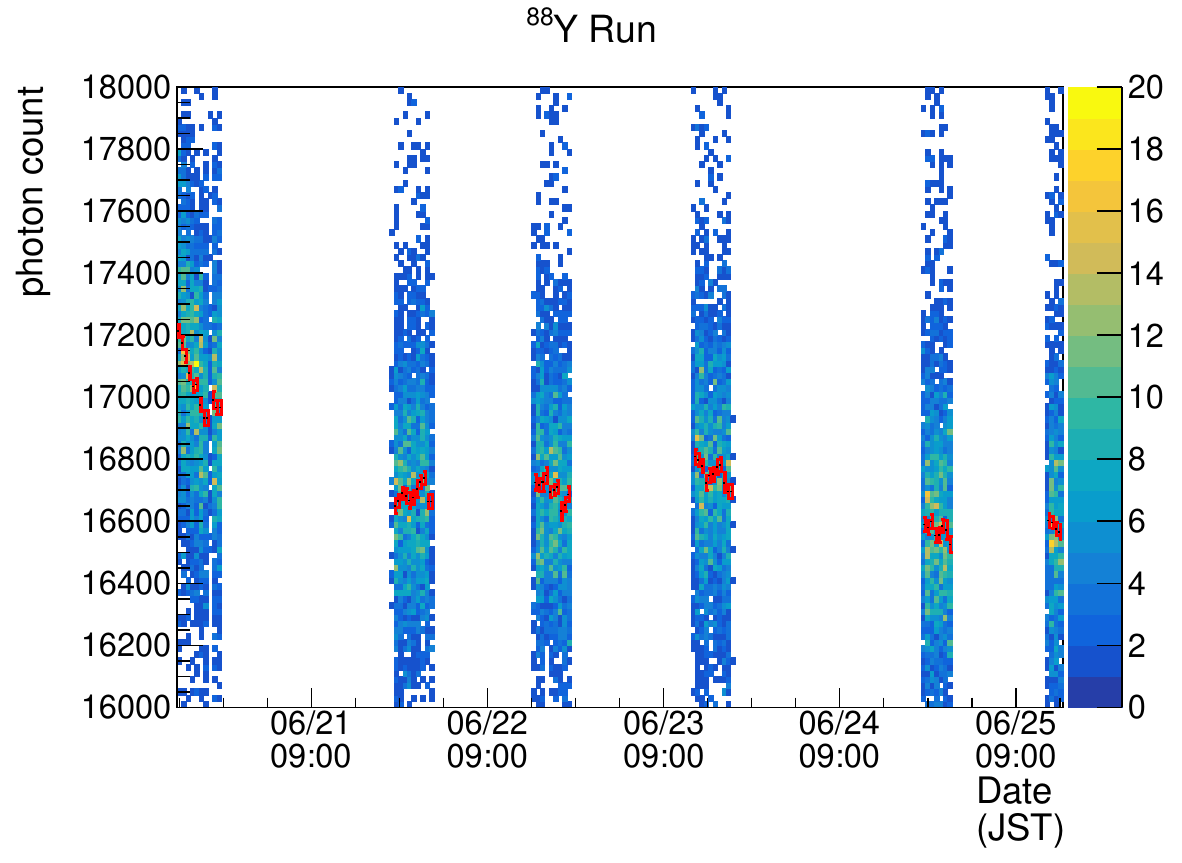}
        \end{minipage}
        \begin{minipage}{0.5\linewidth}
          \centering
          \includegraphics[width=\linewidth]{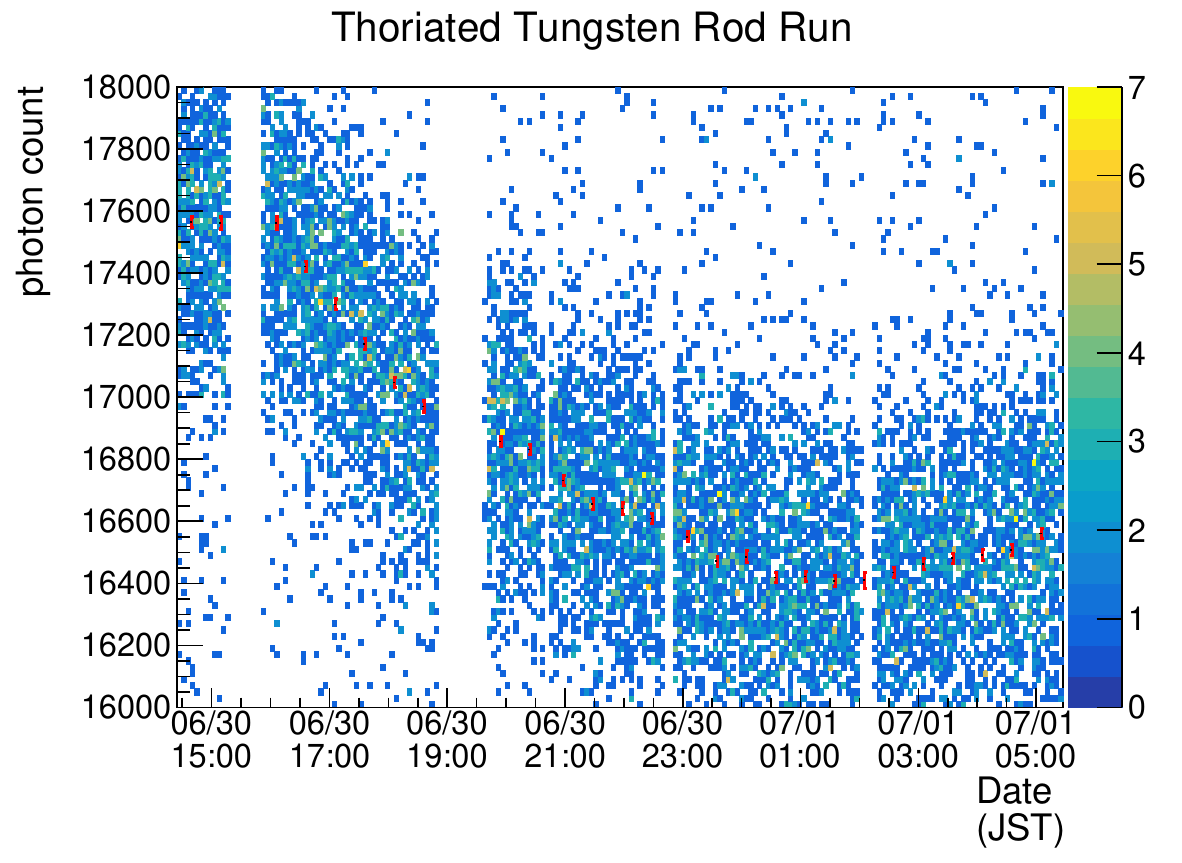}
        \end{minipage}          
        \caption{Variation of photon counts along with the time. The points with the error bar represent the mean of the $\mathrm{K_\alpha}$ peak in each time division. The left panel is for the $^{88}\mathrm{Y}$ run and the right is for the thoriated tungsten rod run.}
        \label{fig:time_dependence}
      \end{figure}

      Figure \ref{fig:time_dependence} shows the time variation of photon counts of clusters around the energy of $\mathrm{K_\alpha}$ characteristic X-ray.
      The cause of variation can be changes in the gas conditions: temperature, density, and purity.
      The correction factors are derived for divisions every 30 minutes.
      The width of the time bin, 30 minutes, is determined so that the width of the peak of \SI{1836}{\keV} is minimized; balanced between the statistical error of the correction factor and the sensitivity to the time variation.
      
    \subsubsection{Correction of $z$-dependence}\label{sec:z_correction}
\begin{figure}[tb]
        \centering
        \includegraphics[width=0.8\linewidth]{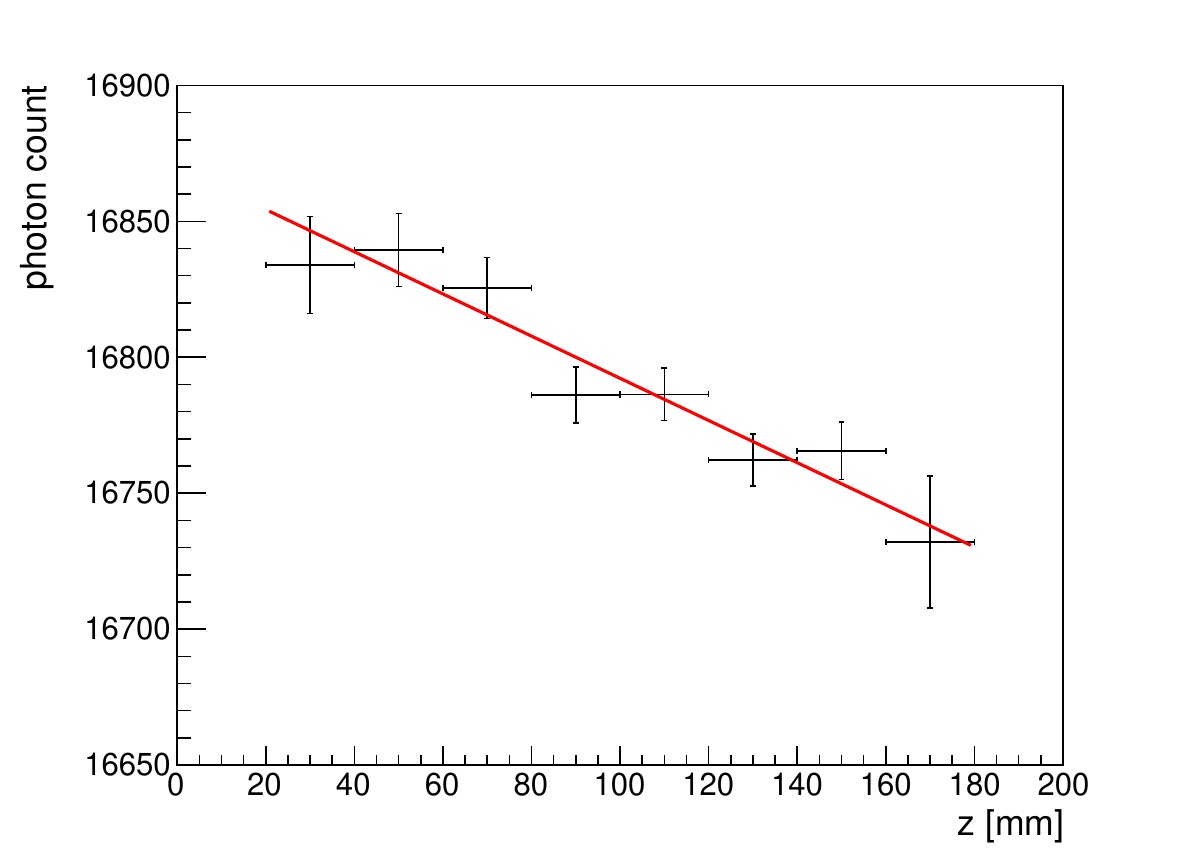}
        \caption{Dependence of the photon counts of $\mathrm{K_\alpha}$ clusters on the $z$-position for the $^{88}\mathrm{Y}$ run. The red line shows the fitted function.}
        \label{fig:z_dependence}
      \end{figure}

      Some amount of the ionization electrons are not detected because of the attachment by impurities during drift.
      This attenuation is characterized by the following equation.
      \begin{equation*}
        N\left(z\right) = N_0\exp\left(-\frac{z}{\lambda}\right) \simeq N_0\left(1-\frac{z}{\lambda}\right),
      \end{equation*}
      where $N_0$ and $N(z)$ are the photon counts before and after the attachment respectively and $\lambda$ is the attenuation length.
      Figure \ref{fig:z_dependence} shows the dependence of the photon counts of $\mathrm{K_\alpha}$ clusters on the $z$-position.
      From this dependence, the attenuation lengths of $\lambda = \SI{21700+-3700}{\mm}$ for the $^{88}\mathrm{Y}$ run and $\lambda = \SI{17000+-2700}{\mm}$ for the thoriated tungsten rod run were obtained.
      These correspond to the electron lifetimes of \SI{20.9+-3.6}{\ms} and \SI{16.3+-2.6}{\ms} respectively.
      Using these attenuation lengths, the photon counts are corrected for every sampling of the waveforms.
      
    \subsubsection{Overall fine-tuning for the recovery times of MPPCs}\label{sec:recovery_time_finetuning}
      As described in sec.\ref{sec:mppc_non_linearity}, the non-linearity of the MPPCs is corrected using the recovery times of MPPCs measured in advance.
      However, the effective recovery times can vary depending on the conditions of the MPPCs, such as temperature, or shadow of the mesh electrode in front of MPPCs.
      The deviation in photon counts due to the difference between the true recovery time and the measured recovery time can be expressed as follows.
      \begin{align}
          \sum_i r^iN_\mathrm{rec}^i - N_\mathrm{true} &= \sum_i \frac{r^iN_\mathrm{obs}^i}{1-k'N_\mathrm{obs}^i} - N_\mathrm{true} \notag\\
                         &\simeq \Delta k\sum_ir^i\left(N_\mathrm{rec}^i\right)^2 ,\label{eq:mppc_finetuning}
      \end{align}
      where $N_\mathrm{true}$ is the true total photon count of the event, $i$ runs for every sampling of the waveform of every hit channel, $N_\mathrm{obs}^i$ and $N_\mathrm{rec}^i$ are the photon count for each sampling of the waveform before and after the MPPC non-linearity correction, respectively, $r^i$ is the correction factor other than the MPPC non-linearity, $k^{\left('\right)} = \tau^{\left('\right)}/\left(\Delta t \cdot N_\mathrm{pixel}\right)$, $\tau^{\left('\right)}$ is the true (measured) recovery time of the channel, and $\Delta k = k - k'$.
      The last line assumes that $\Delta k$ is small and common among channels.
      This equation indicates that, if there exists an overall bias in the recovery times, it appears as a slope of the relation between the photon counts and $\sum_ir^i\left(N_\mathrm{rec}^i\right)^2$ (called hereafter corrected squared sum, CSS).
      
\begin{figure}[tb]
          \centering
          \includegraphics[width=0.8\linewidth]{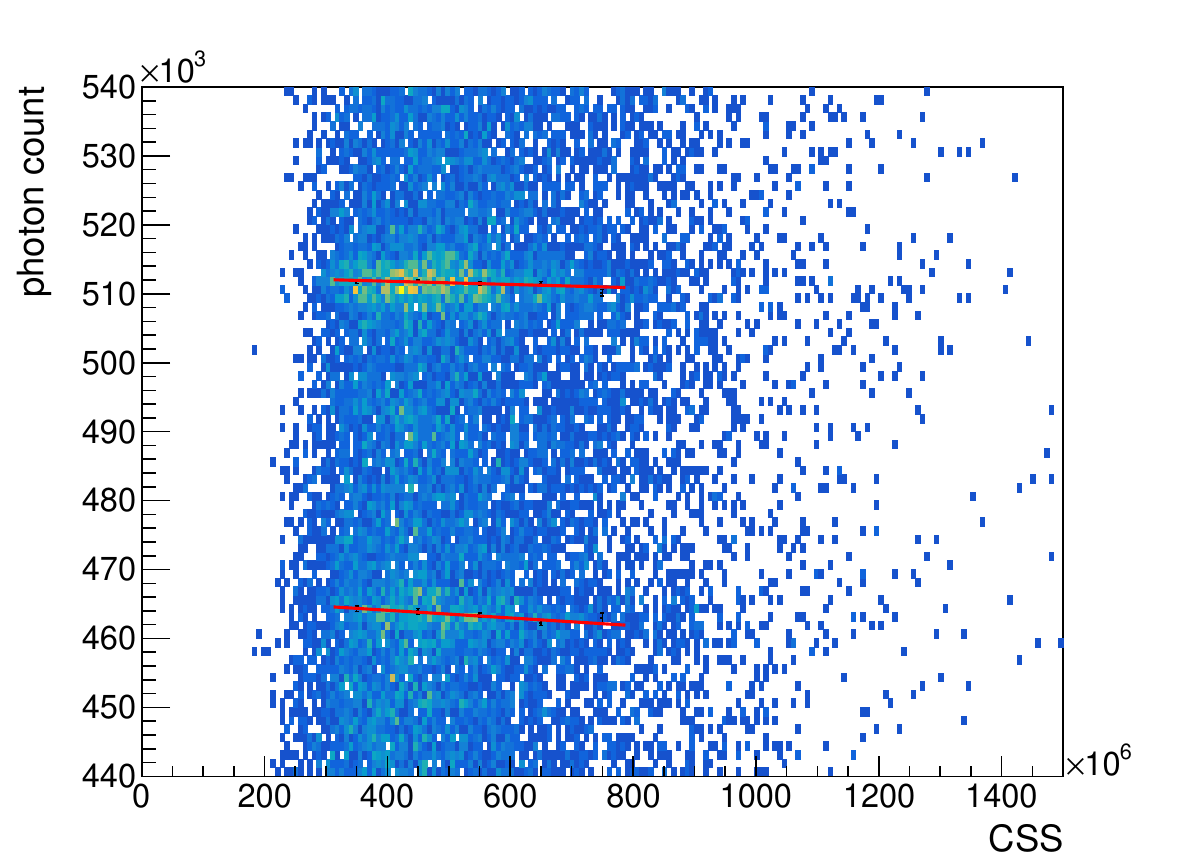}
          \caption{Relation between the photon counts and the CSS for the photopeak of \SI{898}{\keV} gamma rays ($\sim$\num{5.1e5} photon) and the double escape peak of \SI{1836}{\keV} gamma rays ($\sim$\num{4.6e5} photon) in the $^{88}\mathrm{Y}$ run.}
          \label{fig:css}
      \end{figure}

      Figure \ref{fig:css} shows the distribution of the photon counts and the CSS.
      To determine the $\Delta k$ and corresponding biases of the MPPC recovery times, the peaks with sufficient statistics are used; the photopeak of \SI{898}{\keV} gamma rays and the double escape peak of \SI{1836}{\keV} gamma rays for the $^{88}\mathrm{Y}$ run, and the photopeak of \SI{583}{\keV} gamma rays and the double escape peak of \SI{2615}{\keV} gamma rays for the thoriated tungsten rod run.
      From the slope at each peak of the photon counts, the biases for the MPPC recovery times were derived as +\SI{2.35}{\ns} for the $^{88}\mathrm{Y}$ run and +\SI{3.13}{\ns} for the thoriated tungsten rod run.

      The MPPC non-linearity correction, EL gain correction, time variation correction, and $z$ dependence correction are repeated with the recovery times shifted by these biases.

\section{Detector performance}
  From the analysis in the previous section, EL photon count and track of events are obtained.
  Based on these, we evaluate the performance of the detector.
  \subsection{Energy resolution}
\begin{figure}[tb]
    \begin{minipage}{0.5\linewidth}
      \centering
      \includegraphics[width=\linewidth]{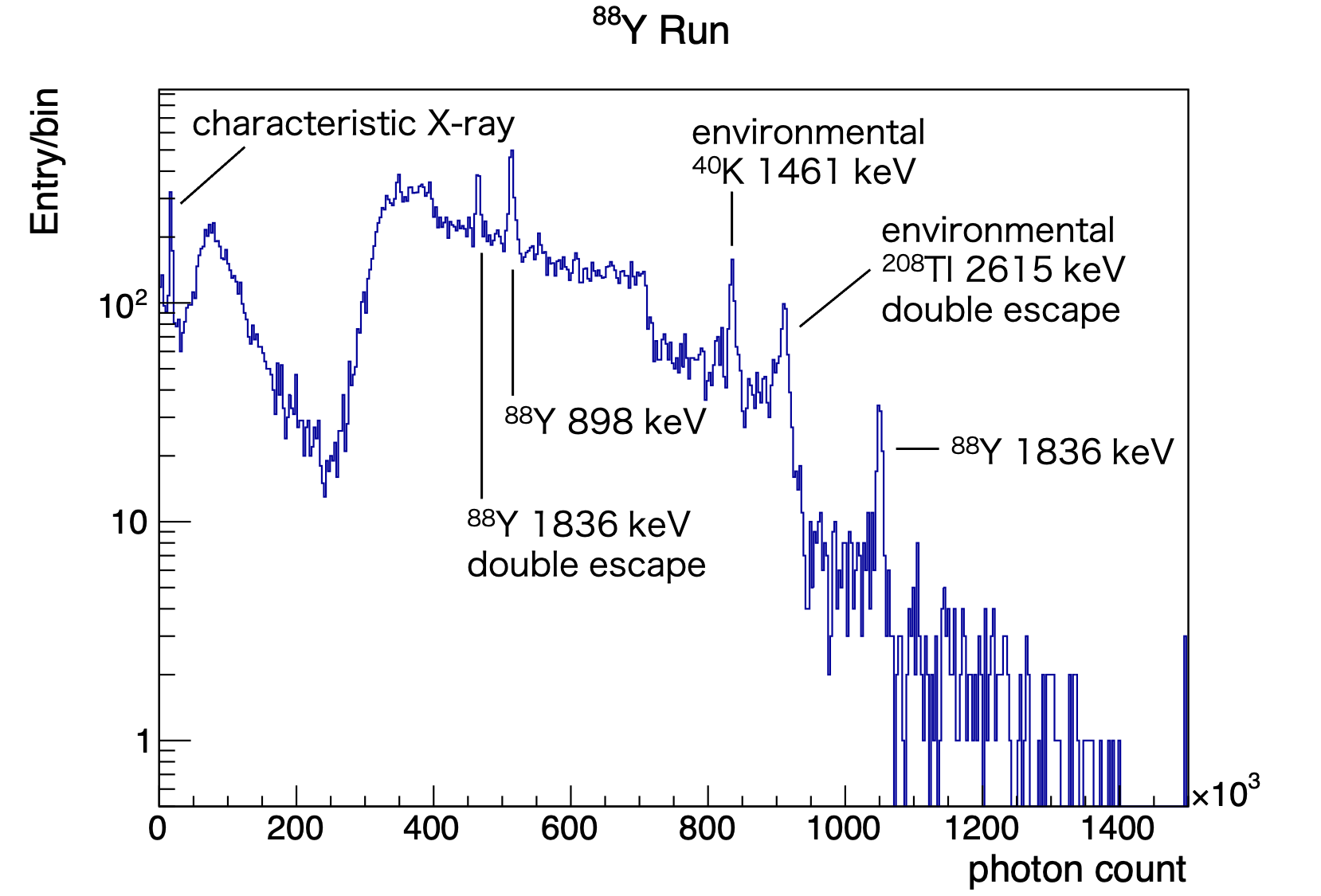}
    \end{minipage}
    \begin{minipage}{0.5\linewidth}
      \centering
      \includegraphics[width=\linewidth]{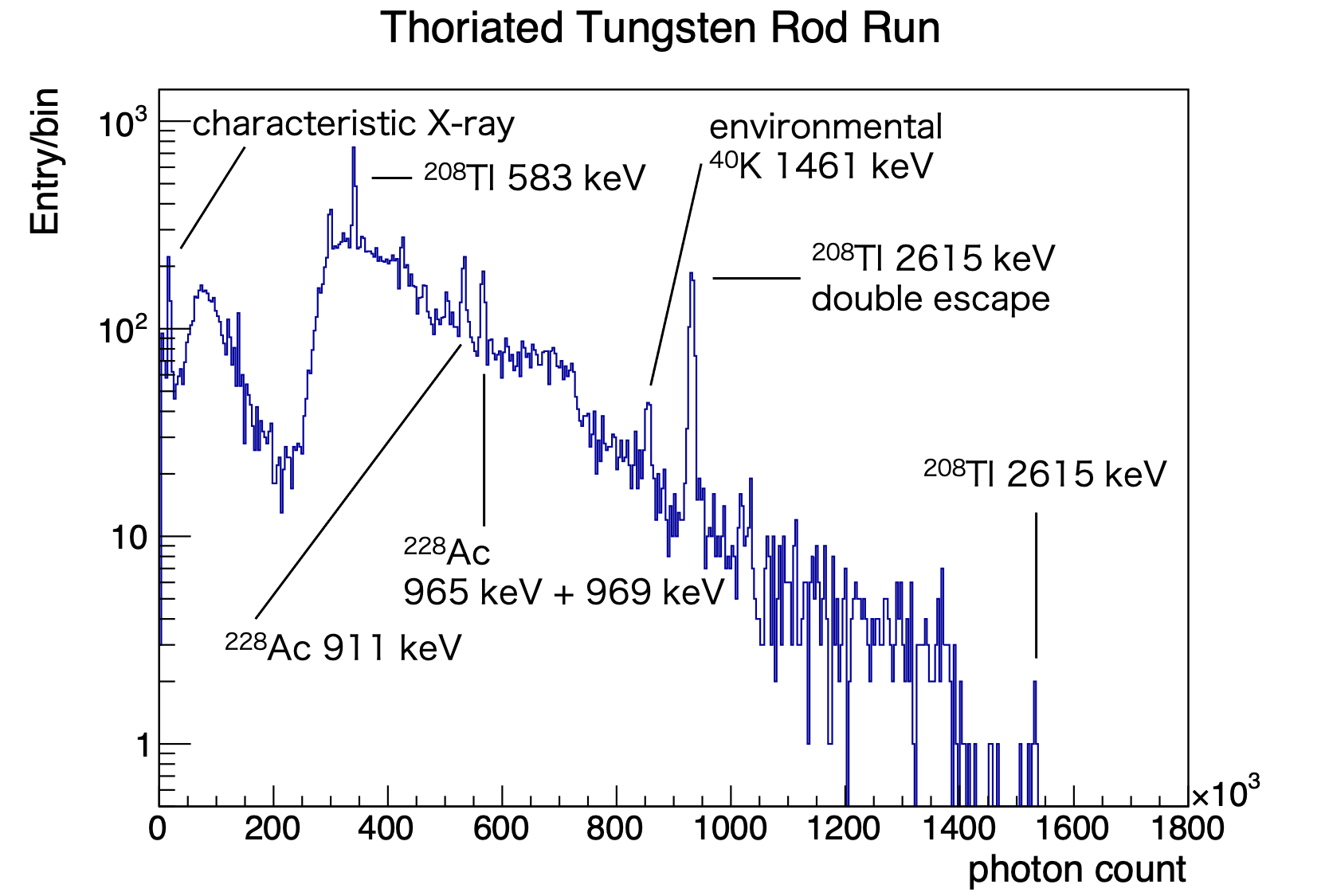}
    \end{minipage}          
    \caption{Photon count spectra after all of the corrections and cuts. The left panel is for the $^{88}\mathrm{Y}$ run and the right is for the thoriated tungsten rod run. The dip around 200 photon count corresponds to the threshold of the fiducial trigger.}
    \label{fig:spectrum}
  \end{figure}

\begin{figure}[tb]
      \centering
      \includegraphics[width=0.8\linewidth]{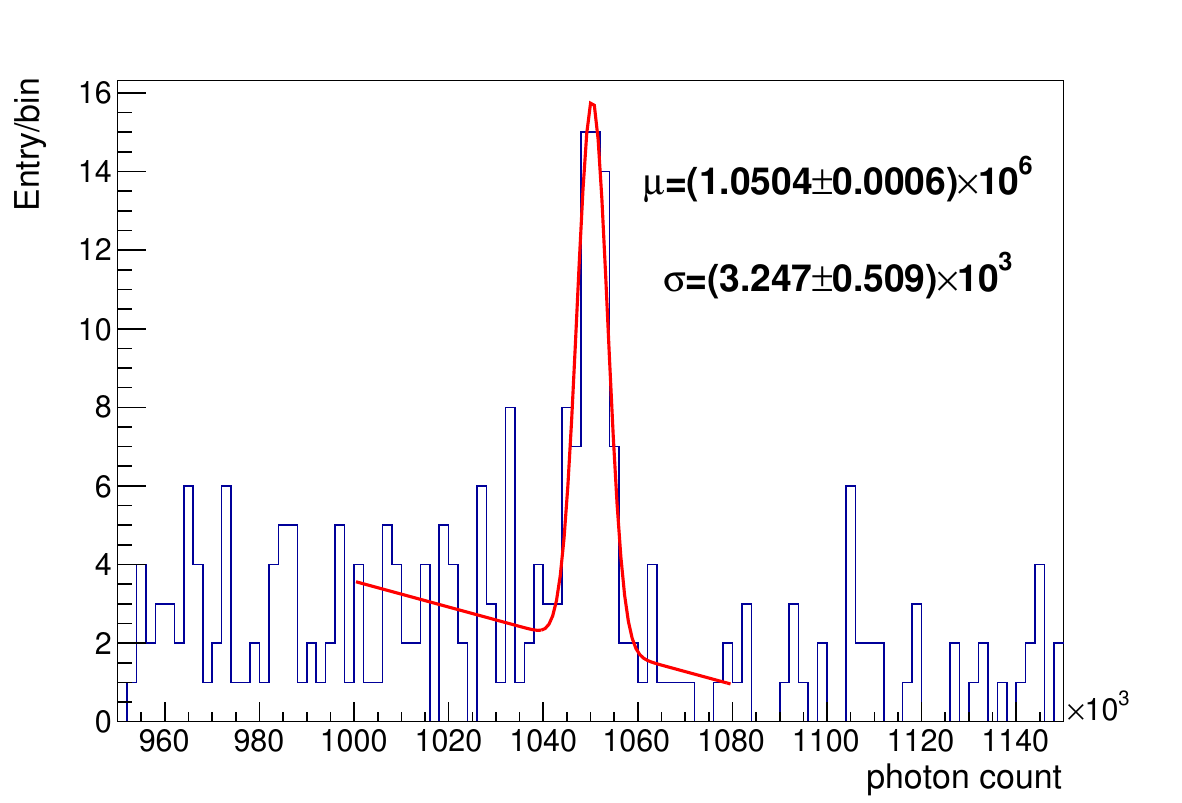}
      \caption{Result of the fit to the spectrum of the single-clustered full energy peak of $^{88}\mathrm{Y}$ \SI{1836}{\keV} gamma rays.}
      \label{fig:fit}
  \end{figure}

  Figure \ref{fig:spectrum} is the EL photon-count spectra of each run.
  Several peaks are identified in the spectra; peaks of characteristic X-rays of xenon, full peaks of gamma rays from the sources and environment, and double escape peaks of pair creation.
  Each peak was fitted assuming a Gaussian peak and linear background.
  For the gamma-ray full peaks, single-cluster events and multi-cluster events were fitted separately.
  Figure \ref{fig:fit} shows an example of the fit results, and Table \ref{tab:fit} is the summary.
  \begin{table}[tb]
      \centering
      \caption{Summary of the result of peak fit. SS stands for the single-site events and MS stands for the multiple-site events for gamma-ray full peaks. $^{40}\mathrm{K}$ multi-cluster events in the thoriated tungsten rod run were too few to evaluate the resolution.}
      \begin{tabular}{c|c|c|c}
       & Energy & mean photon counts & resolution [FWHM] \\
      \hline
      \multicolumn{4}{l}{$^{88}\mathrm{Y}$ run}\\
      \hline
      $\mathrm{K_\alpha}$ & \SI{29.68}{\keV} & \num{1.6870+-0.0004e4} & \SI{4.389+-0.050}{\%} \\
      $\mathrm{K_\beta}$ & \SI{33.62}{\keV} & \num{1.9166+-0.0011e4} & \SI{4.722+-0.125}{\%} \\
      Double escape of $^{88}\mathrm{Y}$ \SI{1836}{\keV} & \SI{814.1}{\keV} & \num{4.6512+-0.0022e5} & \SI{1.194+-0.102}{\%} \\
      $^{88}\mathrm{Y}$ SS & \multirow{2}{*}{\SI{898.0}{\keV}} & \multirow{2}{*}{\num{5.1374+-0.0022e5}} & \SI{1.152+-0.119}{\%} \\
      $^{88}\mathrm{Y}$ MS & & & \SI{1.386+-0.109}{\%} \\
      environmental $^{40}\mathrm{K}$ SS & \multirow{2}{*}{\SI{1461}{\keV}} & \multirow{2}{*}{\num{8.3458+-0.0042e5}} & \SI{0.81+-0.11}{\%} \\
      environmental $^{40}\mathrm{K}$ MS & & & \SI{1.09+-0.16}{\%} \\
      $^{88}\mathrm{Y}$ SS & \multirow{2}{*}{\SI{1836}{\keV}} & \multirow{2}{*}{\num{1.0504+-0.0006e6}} & \SI{0.73+-0.11}{\%} \\
      $^{88}\mathrm{Y}$ MS & & & \SI{0.98+-0.19}{\%} \\
      \hline
      \multicolumn{4}{l}{thoriated tungsten rod run}\\
      \hline
      $\mathrm{K_\alpha}$ & \SI{29.68}{\keV} & \num{1.7270+-0.0005e4} & \SI{4.107+-0.053}{\%} \\
      $\mathrm{K_\beta}$ & \SI{33.62}{\keV} & \num{1.9604+-0.0013e4} & \SI{5.003+-0.155}{\%} \\
      positron annihilation SS & \multirow{2}{*}{\SI{511.0}{\keV}} & \multirow{2}{*}{\num{2.9889+-0.0022e5}} & \SI{1.221+-0.182}{\%} \\
      positron annihilation MS & & & \SI{1.541+-0.362}{\%} \\
      $^{208}\mathrm{Tl}$ SS & \multirow{2}{*}{\SI{583.2}{\keV}} & \multirow{2}{*}{\num{3.4115+-0.0012e5}} & \SI{1.152+-0.078}{\%} \\
      $^{208}\mathrm{Tl}$ MS & & & \SI{1.32+-0.13}{\%} \\
      $^{228}\mathrm{Ac}$ SS & \multirow{2}{*}{\SI{911.2}{\keV}} & \multirow{2}{*}{\num{5.3298+-0.0049e5}} & \SI{1.46+-0.23}{\%} \\
      $^{228}\mathrm{Ac}$ MS & & & \SI{1.17+-0.19}{\%} \\
      environmental $^{40}\mathrm{K}$ SS & \SI{1461}{\keV} & \num{8.5596+-0.0077e5} & \SI{0.65+-0.22}{\%} \\
      Double escape of $^{208}\mathrm{Tl}$ \SI{2615}{\keV} & \SI{1593}{\keV} & \num{9.3178+-0.0020e5} & \SI{0.940+-0.044}{\%} \\
      \end{tabular}
      \label{tab:fit}
  \end{table}
  
\begin{figure}[tb]
    \begin{tabular}{cc}
      \begin{minipage}{0.5\linewidth}
        \centering
        \includegraphics[width=\linewidth]{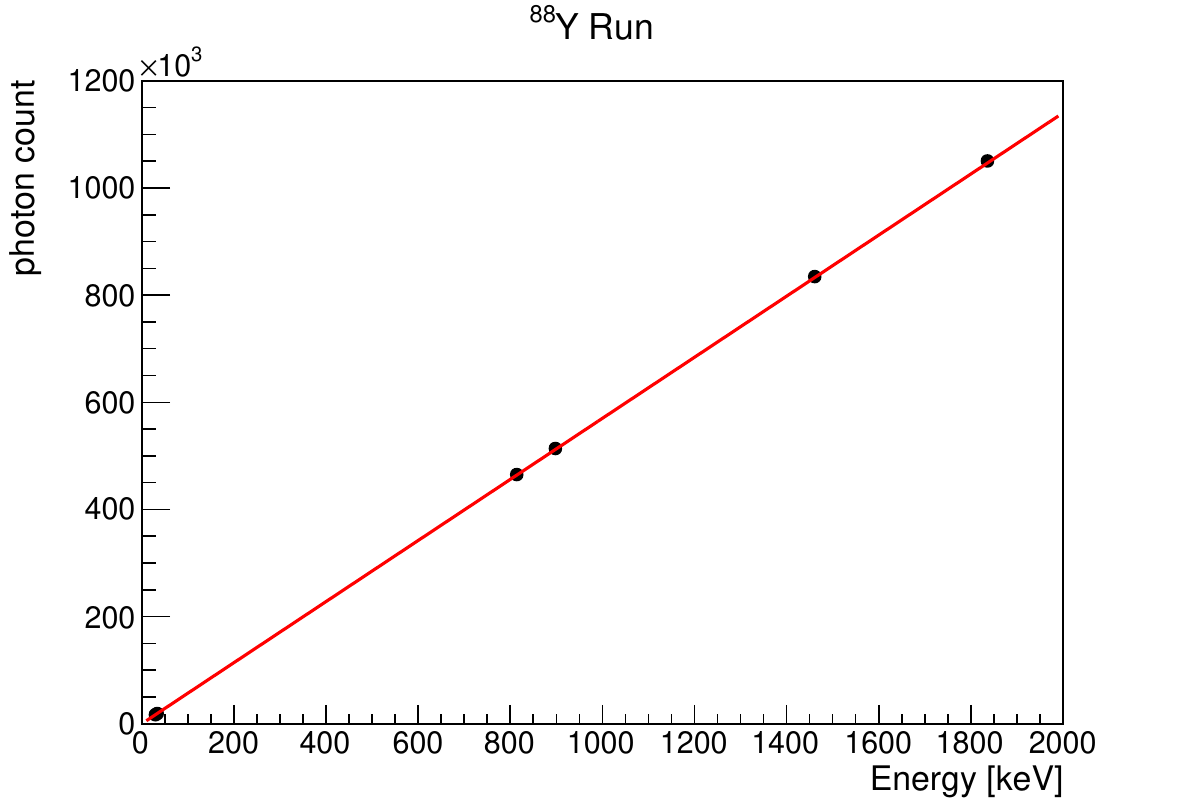}
        \subcaption{}
        \label{fig:linearity_88Y}
      \end{minipage} &
      \begin{minipage}{0.5\linewidth}
        \centering
        \includegraphics[width=\linewidth]{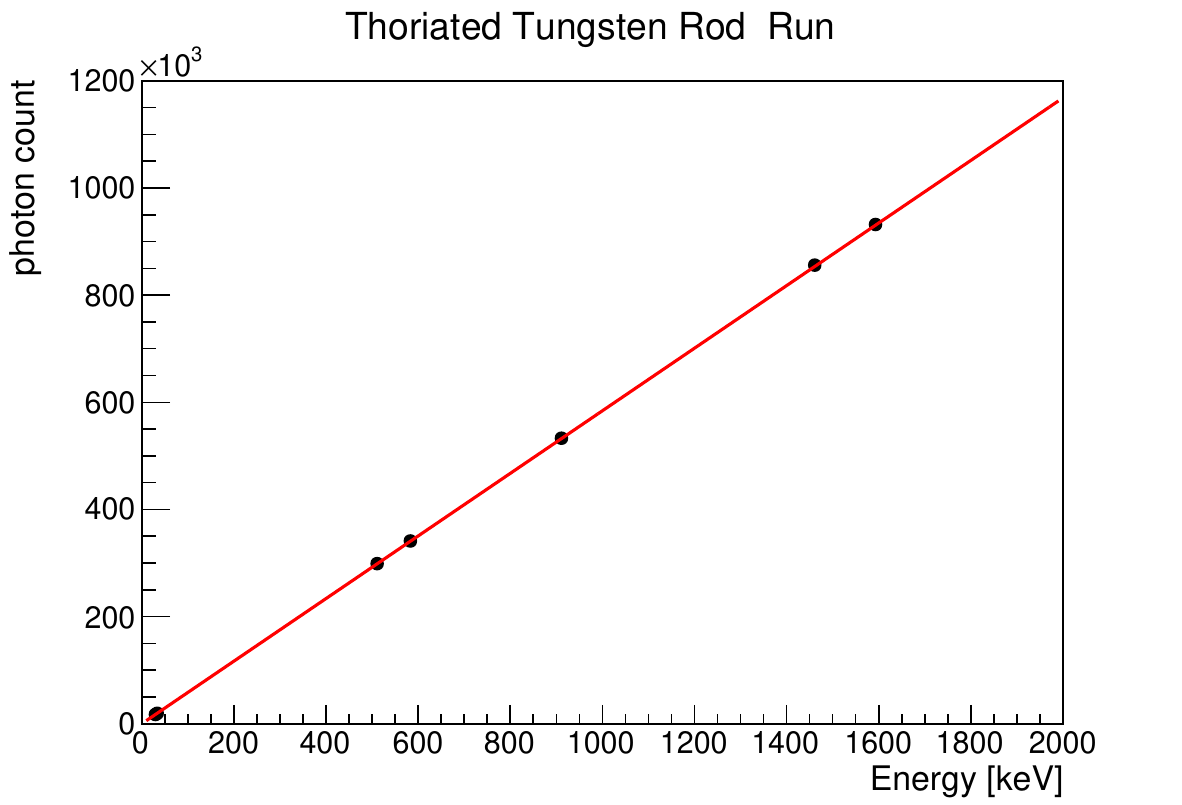}
        \subcaption{}
        \label{fig:linearity_tt}
      \end{minipage}\\
      \begin{minipage}{0.5\linewidth}
        \centering
        \includegraphics[width=\linewidth]{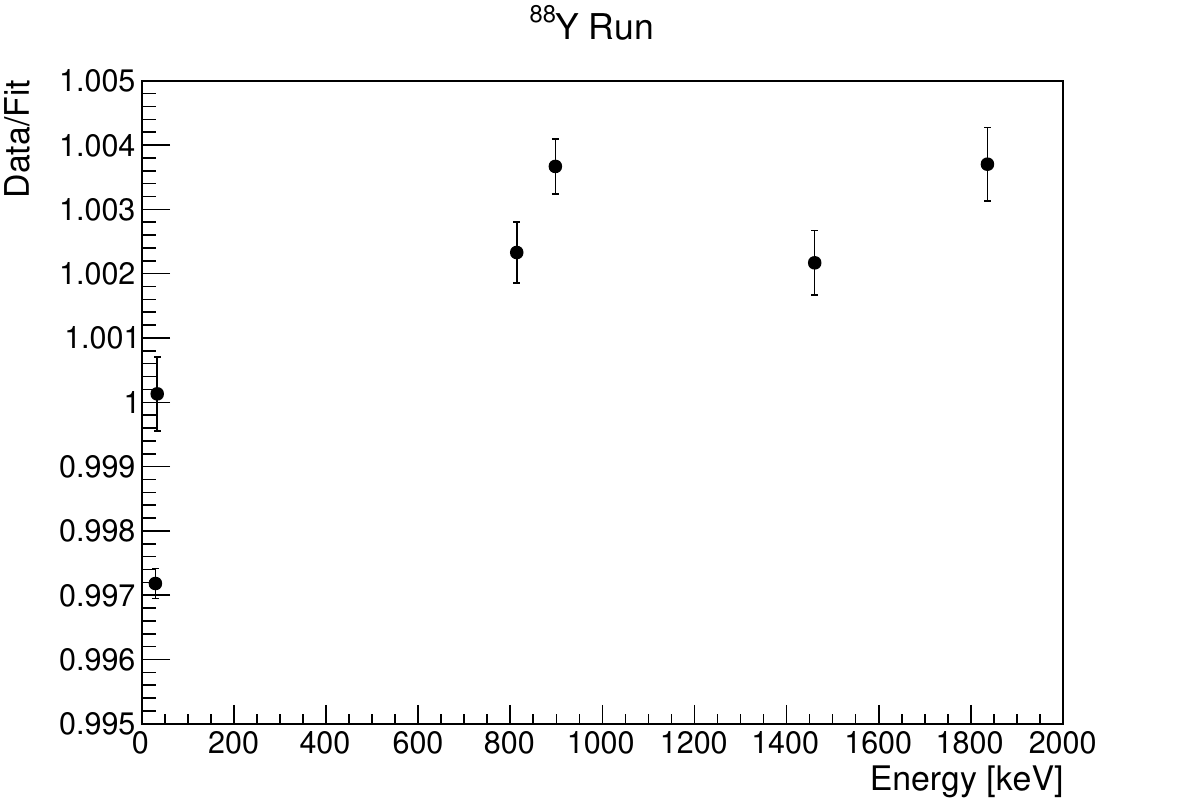}
        \subcaption{}
        \label{fig:linearity_ratio_88Y}
      \end{minipage} &
      \begin{minipage}{0.5\linewidth}
        \centering
        \includegraphics[width=\linewidth]{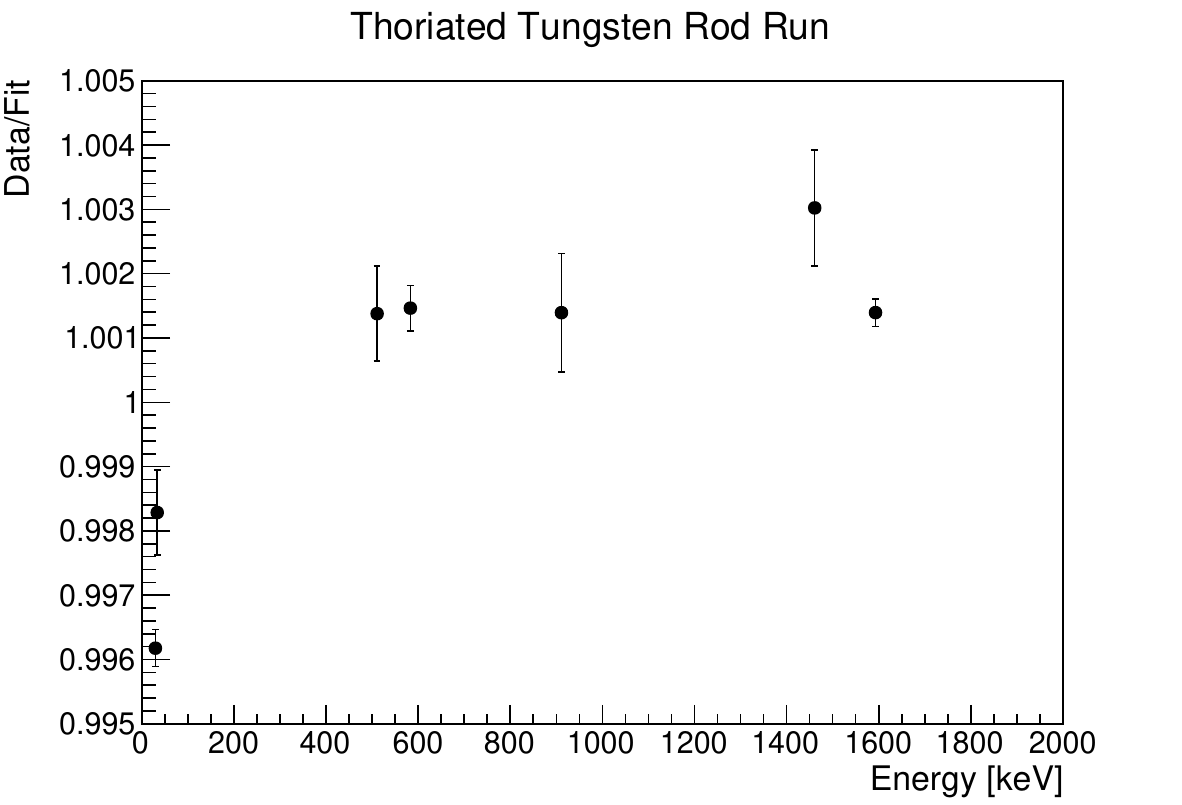}
        \subcaption{}
        \label{fig:linearity_ratio_tt}
      \end{minipage}
    \end{tabular}
    \caption{Relation between the photon counts and the corresponding energies. The lines are the fit results as proportional ((a) and (b)). The ratio of the data point to the fit ((c) and (d)).}
  \end{figure}

  Figures \ref{fig:linearity_88Y} and \ref{fig:linearity_tt} show the mean photon counts of each peak versus the energy from \cite{NuclearDataSearch}.
  The ratio of the data point to the fitted proportional line is shown in Figs. \ref{fig:linearity_ratio_88Y} and \ref{fig:linearity_ratio_tt}.
  Linearity is good except that the $\mathrm{K_\alpha}$ peaks are below the fitted line.

\begin{figure}[tb]
      \centering
      \includegraphics[width=0.8\linewidth]{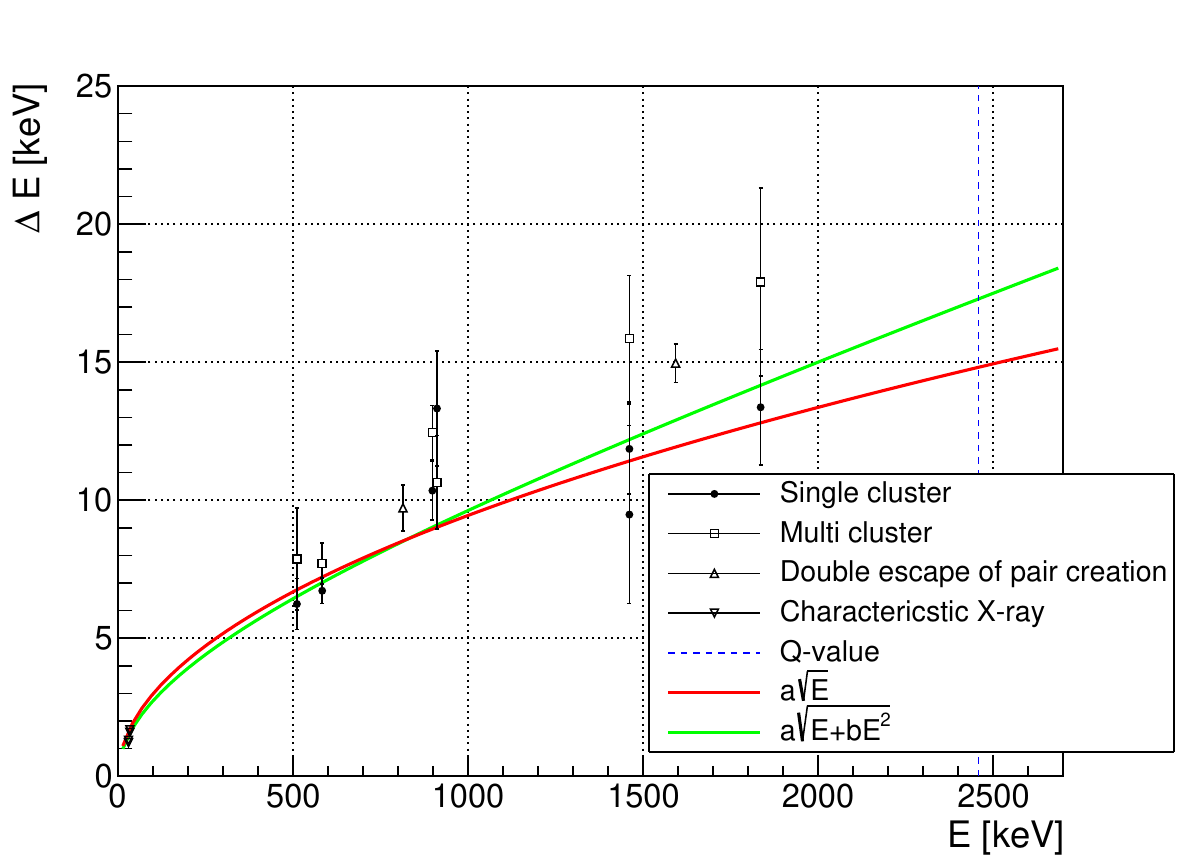}
      \caption{Extrapolation of the energy resolution to the Q value of $^{136}\mathrm{Xe}$ $0\nu\beta\beta$ with two kinds of the fit function, $a\sqrt{E}$ and $a\sqrt{E+bE^2}$. Only the single-cluster gamma-ray data points (solid circle) were used for the fit.}
      \label{fig:resolution}
  \end{figure}

  By extrapolating these results, we estimate the energy resolution at the Q value of $^{136}\mathrm{Xe}$ $0\nu\beta\beta$, \SI{2458}{\keV}.
  Two cases are considered for the dependence on $E$; $a\sqrt{E}$ and $a\sqrt{E+bE^2}$.
  The former is for a situation dominated by statistical fluctuation, and the latter is with systematics contributing.
  Figure \ref{fig:resolution} shows the results of the extrapolation to the Q value.
  Note that only the data points of single-cluster gamma-ray peaks are used.
  The estimated energy resolution at the Q value is \SI{0.60+-0.03}{\%} for the form of $a\sqrt{E}$ and \SI{0.70+-0.21}{\%} for the form of $a\sqrt{E+bE^2}$.
  The multiple-clustered events give slightly worse resolutions.
  Possible reasons are discussed in Sec.~\ref{sec:summary_resolution}.

  \subsection{Breakdown of the energy resolution}
  Contributions from various sources to the energy resolution were evaluated for the peak of $^{88}\mathrm{Y}$ \SI{1836}{\keV} gamma-ray as follows.
  
    \subsubsection{Fluctuation in the signal generation process}
    The following five factors are considered in this category; fluctuation of the number of initial ionization electrons, recombination, attachment, fluctuation of the EL generation, and fluctuation of the MPPC non-linearity.
    
    The fluctuation of the number of initial ionization electrons is calculated to be 0.29\% with Fano factor of 0.13\cite{ANDERSON1979125} and the W-value of \SI{22.1}{\eV}\cite{RevModPhys.52.121}.

    The energy resolution deteriorates as the drift electric field is lowered.
    This is because of the recombination of ionization electrons.
    The energy resolution for \SI{661.7}{\keV} gamma rays is 0.6\% at $\gtrsim$\SI{100}{\V/\cm/bar} but is worsened to 0.7\% at the electric field at which we performed our measurement (\SI{83.3}{\V/\cm/bar})\cite{BOLOTNIKOV1997360}.
    This difference corresponds to 0.22\% at 1836 keV.

    The number of ionization electrons is reduced by 0.83\% by attachment during the \SI{180}{\mm} drift with the measured attenuation length of \SI{21700}{\mm}.
    Then, the fluctuation of this reduction is at most 0.02\%.
    
    The fluctuation of the EL generation and detection was evaluated by a simulation tuned with the measured EL gain, and was found to be 0.24\%.

    As discussed in Sec.~\ref{sec:mppc_non_linearity}, MPPCs suffer from non-linearity when the number of photons simultaneously incident is close to the number of pixels.
    This is a statistical process so the fluctuation remains even after the non-linearity is corrected.
    By comparing the simulations with and without this effect, the contribution was estimated to be 0.18\%.
    
    \subsubsection{Calibration error}
    Errors in the following four corrections can contribute to the energy resolution; EL gain correction, MPPC recovery times, time variation correction, and $z$-dependence correction.

    The contribution from the error of the EL gain correction (Sec. \ref{sec:elgain}) is calculated as follows,
    \begin{equation*}
      \frac{\sqrt{\sum_\mathrm{ch}\left(\epsilon_\mathrm{ch}\bar N_\mathrm{ch}\right)^2}}{\bar{N}}\times2.36 \simeq \bar{\epsilon}\sqrt{\frac{\sum_\mathrm{ch}\bar{N}_\mathrm{ch}^2}{\bar{N}^2}}\times2.36 ,
    \end{equation*}
    where $\epsilon_\mathrm{ch}$ is the error for each channel, $\bar{\epsilon}$ is the mean error, $\bar{N}_\mathrm{ch}$ is the mean photon count for each channel, $\bar{N}$ is the mean total photon count at \SI{1836}{keV}, and 2.36 is the conversion factor from the standard deviation to the FWHM.
    As $\bar{\epsilon} = 0.46\%$ and ${\sum_\mathrm{ch}\bar{N}_\mathrm{ch}^2}/{\bar{N}^2} = 0.043$, the contribution to the energy resolution is 0.23\%.
    This result is also interpreted as $\bar{\epsilon}/\sqrt{n_\mathrm{eff}}\times2.36$, where $n_\mathrm{eff} = 22.7$ is the effective number of the hit channels.

    The accuracy of the MPPC recovery time measurement affects the energy resolution in two ways: precision of individual MPPC's recovery times and overall bias.
    The recovery times of individual MPPCs were measured with about \SI{0.5}{\ns} precision. Its effect was estimated by simulation and found to be negligible.
    The effect of the overall bias is evaluated based on the Eq.(\ref{eq:mppc_finetuning}) in Sec.~\ref{sec:recovery_time_finetuning}.
    After the overall fine-tuning of the recovery times, $\Delta k$ is \num{-0.29+-1.84e-6}, consistent with zero, which is thanks to the fine-tuning.
    For \SI{1836}{\keV} events, the FWHM of the distribution of CSS is \num{6.15e8}, and therefore the contribution to the energy resolution is at most $\sqrt{0.29^2+1.84^2}\times10^{-6}\times6.15\times10^8/\left(1.05\times10^6\right) = 0.11\%$.

    The time variation correction factor is determined from the $\mathrm{K_\alpha}$ peak fit in each time bin (Sec. \ref{sec:time_correction}).
    The average fit error is 0.137\%, therefore the error of the scale factor is also 0.137\%, and the contribution to the energy resolution is 0.32\%, multiplied by 2.36.
    
    The variation within the time bin is also evaluated.
    There is at most 0.24\% variation in the time bin of 30 minutes.
    Assuming the variation in the time bin is uniform, the contribution to the energy resolution is at most $0.24\%\times\frac{2.36}{\sqrt{12}}=0.16\%$.
    
\begin{figure}[tb]
      \centering
      \includegraphics[width=0.8\linewidth]{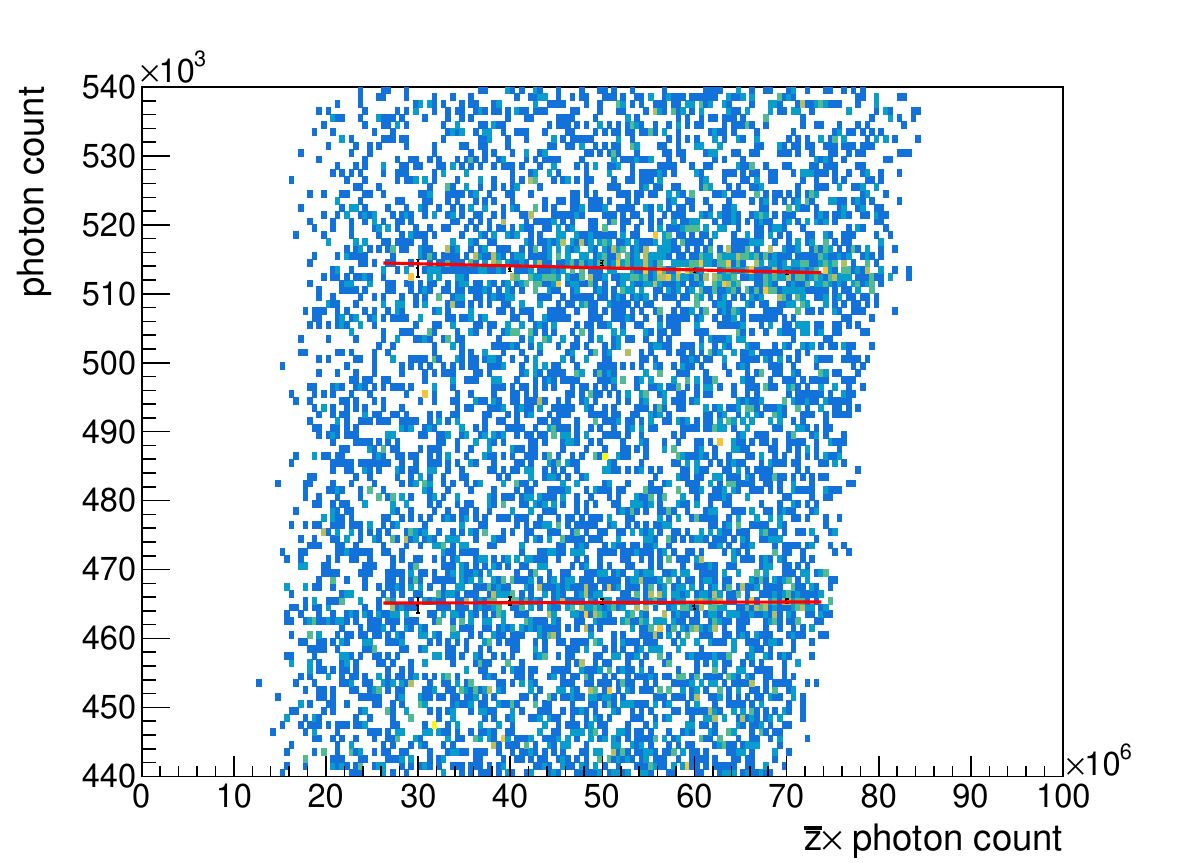}
      \caption{Relation between the corrected photon counts ($N_\mathrm{cor}$) and the product of the photon counts and the mean $z$-posision ($\bar{z}N_\mathrm{cor}$). The clusters at $\sim$\num{5.1e5} photons and $\sim$\num{4.6e5} photons correspond to the photopeak of 898 keV gamma rays and the double escape peak of 1836 keV gamma rays, respectively.}
      \label{fig:z_cm}
    \end{figure}

    When there is an error in the attenuation length determination, the error from the $z$-correction (Sec.\ref{sec:z_correction}) on the photon counts is calculated as
    \begin{align*}
      \Delta N_\mathrm{cor} &= \sum_ip^iN_\mathrm{obs}^i\left(1+\frac{z^i}{\lambda'}\right)-\sum_ip^iN_\mathrm{obs}^i\left(1+\frac{z^i}{\lambda}\right) \\
                     &\simeq \left(\frac{1}{\lambda'}-\frac{1}{\lambda}\right)\bar{z}N_\mathrm{cor}
    \end{align*}
    where $p^i$ is the correction factor other than the $z$-dependence correction, $z^i$ is the $z$-position of each sampling of the waveform, $\lambda'$ is the attenuation length used in the correction, $\lambda$ is the true attenuation length, and $\bar{z}$ is the $z$-position of the event given as the mean weighted by the photon counts.
    Figure \ref{fig:z_cm} shows the distribution of $N_\mathrm{cor}$ versus $\bar{z}N_\mathrm{cor}$.
    $1/{\lambda'} - 1/\lambda$ would appear as a slope of clusters.
    From this plot, $1/{\lambda'} - 1/\lambda$ is obtained as \SI{-1.52+-1.12e-5}{\per\mm}, consistent with zero.
    Because the FWHM of the distribution of $\bar{z}N_\mathrm{cor}$ for \SI{1836}{\keV} events is \SI{6.03e7}{\mm}, the contribution to the energy resolution is at most $\sqrt{1.52^2+1.12^2}\times{10}^{-5}\times6.03\times{10}^{7}$ photons, that is 0.11\%.

    \subsubsection{Hardware-origin error}
    Position dependence of the EL gain and errors arising from the waveform processing in the FEB are considered.

    The EL gains depend on the injection positions of ionization electrons relative to the cell.
    The amount of the dependence was estimated from the calculated electric field distribution\cite{ban2020} and the effect on the energy resolution was found to be negligible by comparing the results of simulations with and without this dependence.

    In the FEBs, the signal waveforms are shaped by Sallen-Key filters and then digitized.
    The effect of these filtering and digitization was evaluated by simulation and was found to be negligible.
    
    The baseline of the waveform is unknown within one ADC count.
    This leads to two effects on the energy reconstruction.
    First, event-by-event fluctuation of the unknown offset causes fluctuation in the photon count determination.
    In addition, since the event time width itself fluctuates, the offset affects the photon count determination even if it is constant.
    The contribution to the energy resolution from the baseline offset is calculated from the mean and standard deviation of the event time width and was found to be 0.09\% at most.
    The contributions from hardware are small.
    This is natural because they were so designed.
    
    \subsubsection{Mis-reconstruction of $z$-position}
\begin{figure}[tb]
      \centering
      \includegraphics[width=0.8\linewidth]{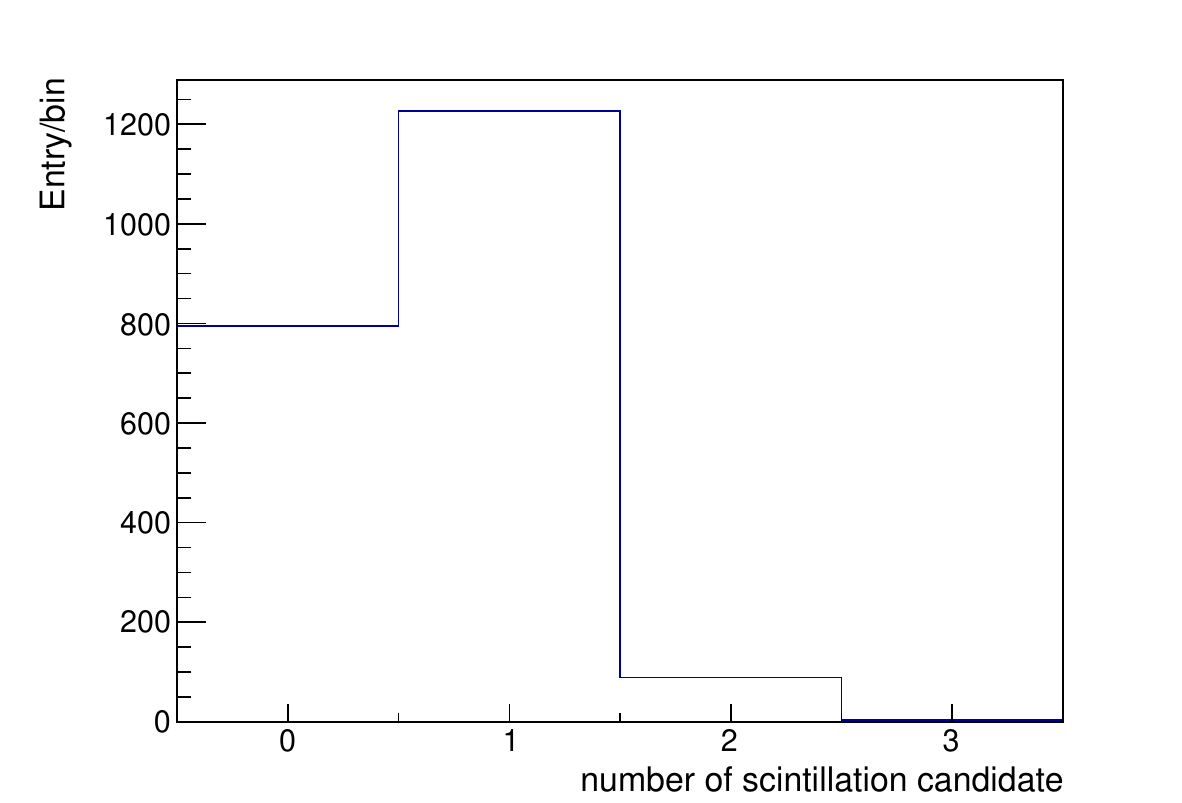}
      \caption{Number of candidates for the primary scintillation light for \SI{1836}{\keV} events.}
      \label{fig:num_sci_candidate}
    \end{figure}

    If the primary scintillation is wrongly identified, the $z$-position of the event is mis-reconstructed and the correction of $z$-dependence is wrongly applied.
    Figure \ref{fig:num_sci_candidate} shows the distribution of the number of hit clusters for the primary scintillation light candidates for \SI{1836}{\keV} events.
    Assuming the efficiency of detecting the right scintillation light is $\varepsilon$, and the average number of the detected accidental scintillation hits is $\mu_\mathrm{acc}$, the probabilities that no or just one hit cluster is detected as a primary scintillation light candidate are as follows.
    \begin{align*}
      P\left(n_\mathrm{sci}=0\right) &= \left(1-\varepsilon\right)e^{-\mu_\mathrm{acc}}\\
      P\left(n_\mathrm{sci}=1\right) &= \left(1-\varepsilon\right)\mu_\mathrm{acc}e^{-\mu_\mathrm{acc}} + \varepsilon e^{-\mu_\mathrm{acc}}
    \end{align*}
    From Fig. \ref{fig:num_sci_candidate}, it follows that $ \varepsilon = 0.60$ and $\mu_\mathrm{acc} = 0.075$.
    Since only the events with just one hit cluster are chosen in the analysis, the probability of mis-reconstruction of the $z$-position is $\frac{\left(1-\varepsilon\right)\mu_\mathrm{acc}e^{-\mu_\mathrm{acc}}}{P\left(n_\mathrm{sci}=1\right)}=5\%$.
    Assuming the mis-reconstruction distributes uniformly from \SI{0}{\mm} to \SI{180}{\mm}, the mis-correction of $z$-dependence uniformly distributes from 0 to 0.83\%.
    Then, the contribution to the energy resolution is $\sqrt{5\%}\times\frac{0.83\%}{\sqrt{12}}\times2.36=0.13\%$.

    \subsubsection{Summary of the energy resolution breakdown and prospect of improvement}\label{sec:summary_resolution}
    Table \ref{tab:resolution_breakdown} summarizes the breakdown of the energy resolution at \SI{1836}{\keV}.
    \begin{table}[tb]
      \centering
      \caption{Breakdown of the energy resolution at \SI{1836}{\keV} listed in descending order.}
      \begin{tabular}{c|c}
        Error in the time variation correction                    & 0.32 \% \\
        Fluctuation of the number of initial ionization electrons & 0.29 \% \\
        Fluctuation of the EL generation and detection            & 0.24 \% \\
        Error in the EL gain correction                           & 0.23 \% \\
        Recombination                                             & 0.22 \% \\
        Fluctuation of the MPPC non-linearity                     & 0.18 \% \\
        $z$ mis-reconstruction                                    & 0.13 \% \\
        Variation in time bin of time variation correction        & $\lesssim$ 0.16 \% \\
        Error in the $z$-dependence correction                    & $\lesssim$ 0.11 \% \\
        Accuracy of the MPPC recovery times                       & $\lesssim$ 0.11 \% \\
        Offset of the baseline                                    & $\lesssim$ 0.09 \% \\
        Fluctuation of the attachment                             & $\lesssim$ 0.02 \% \\
        Position dependence of the EL gain                        & 0 \% \\
        Waveform processing in the FEB                            & 0 \% \\
        \hline
        Estimation total                                          & \SIrange{0.63}{0.67}{\%} \\
        \hline
        Data total                                                & \SI{0.73+-0.11}{\%}
      \end{tabular}
      \label{tab:resolution_breakdown}
    \end{table}
    The total estimated energy resolution is \SIrange{0.63}{0.67}{\%} while the measured energy resolution is \SI{0.73+-0.11}{\%}. They are in agreement within the margin of error.
    The estimation was made for the single-clustered track case.
    We figure that the worse energy resolution for the multiple-clustered events is because of larger contributions from recombination, fluctuation of the MPPC non-linearity, and accuracy of the MPPC recovery times.
    
    The fluctuation of the EL generation and detection can be suppressed by increasing the detected number of photons.
    We are developing new ELCC with MPPCs of approximately two times larger sensitive areas and anode electrodes with higher discharge resistance.
    Recombination can be suppressed by applying a stronger drift electric field, which is now limited by the discharges at ELCC.
    The accuracies of the EL gain correction and the time variation correction are limited by the statistics of the $\mathrm{K}_\alpha$ peak events and therefore can be reduced by taking more data with a steadier condition.
    Mis-reconstruction of $z$-position comes from the limited efficiency of the primary scintillation detection, which is now 1 p.e. level.
    We are developing a wavelength-shifting-plate configuration to improve the efficiency of primary scintillation detection and reduce mis-reconstruction using the information of photon counts.
    With these countermeasures, the total energy resolution is expected to be improved down to 0.37\% (FWHM) at \SI{1836}{\keV}, which corresponds to 0.32\% (FWHM) at the Q value.
  
  \subsection{Track reconstruction}
\begin{figure}[tb]
      \centering
      \includegraphics[width=0.95\linewidth]{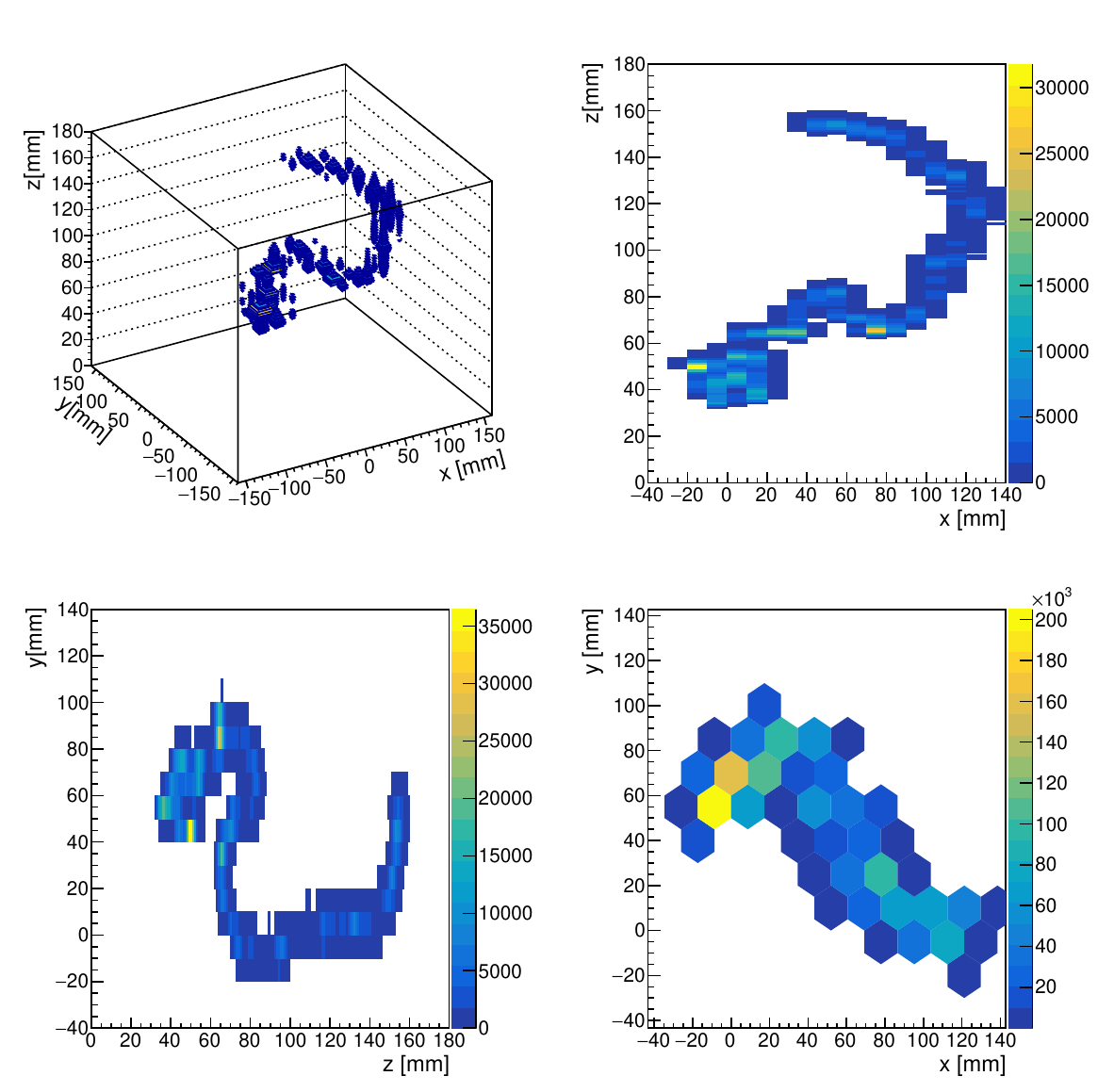}
      \caption{Reconstructed track image of a \SI{2615}{\keV} event. It is considered to be a photoabsorption event.}
      \label{fig:2615_keV_photoabsorption}
  \end{figure}

\begin{figure}[tb]
      \centering
      \includegraphics[width=0.95\linewidth]{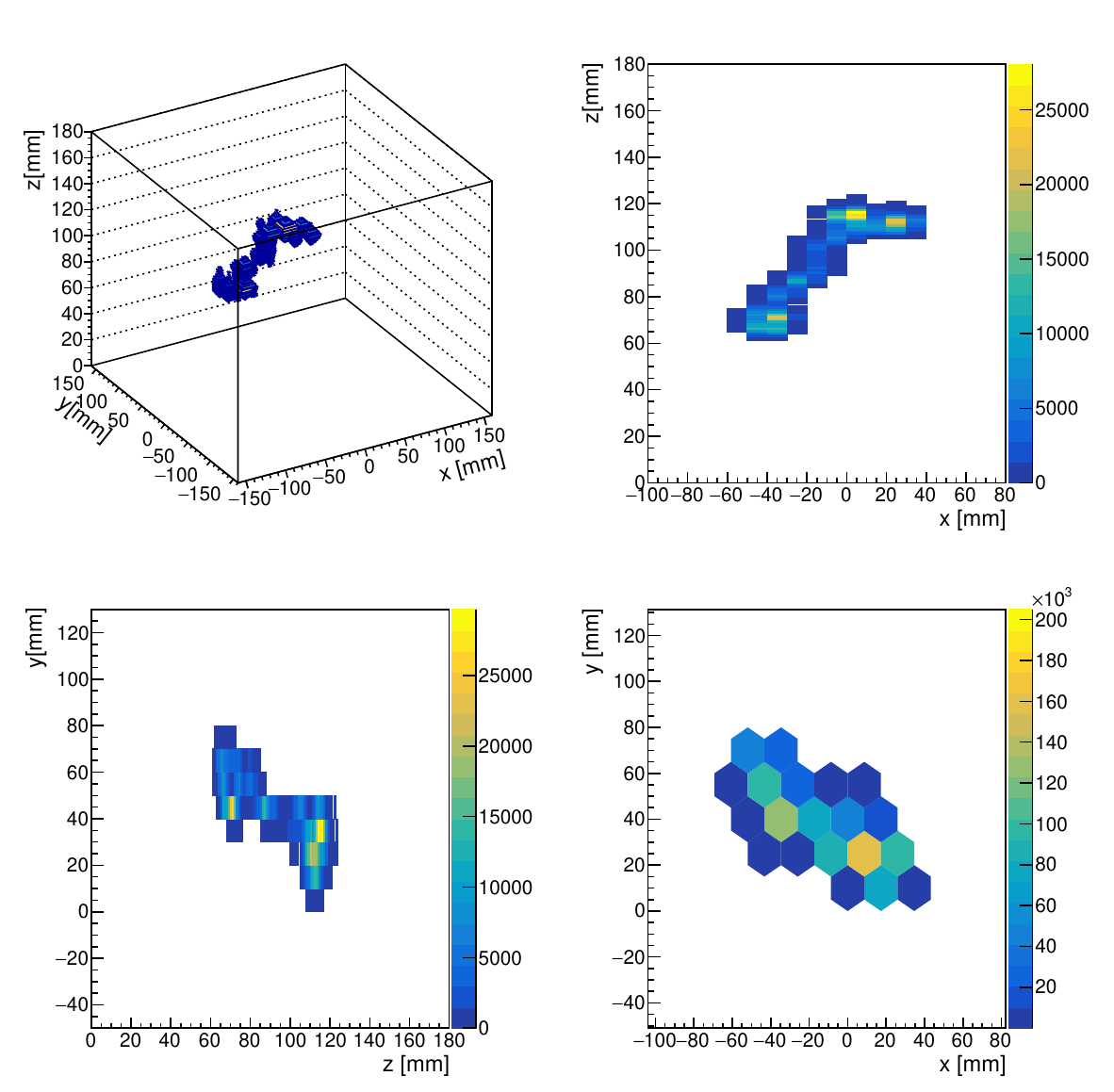}
      \caption{Reconstructed track image of a \SI{1593}{\keV} event. It is considered to be a double escape event of a pair creation by a \SI{2615}{\keV} gamma ray.}
      \label{fig:pair_creation}
  \end{figure}

  Figures \ref{fig:2615_keV_photoabsorption} and \ref{fig:pair_creation} are typical reconstructed track images of a \SI{2615}{\keV} event and a \SI{1593}{\keV} event.
  The former is consistent with a photoabsorption of a \SI{2615}{\keV} gamma ray from $^{208}\mathrm{Tl}$.
  The latter is considered to be a double escape of a \SI{2615}{\keV} pair creation.
  A dense energy deposit at the end of the track ("blob") can be seen in Fig.~\ref{fig:2615_keV_photoabsorption} and two blobs in Fig.~\ref{fig:pair_creation}.
  The number of blobs will be a key to distinguishing the $0\nu\beta\beta$ signals from the gamma-ray backgrounds.
  
\begin{figure}[tb]
        \begin{minipage}{0.5\linewidth}
          \centering
          \includegraphics[width=\linewidth]{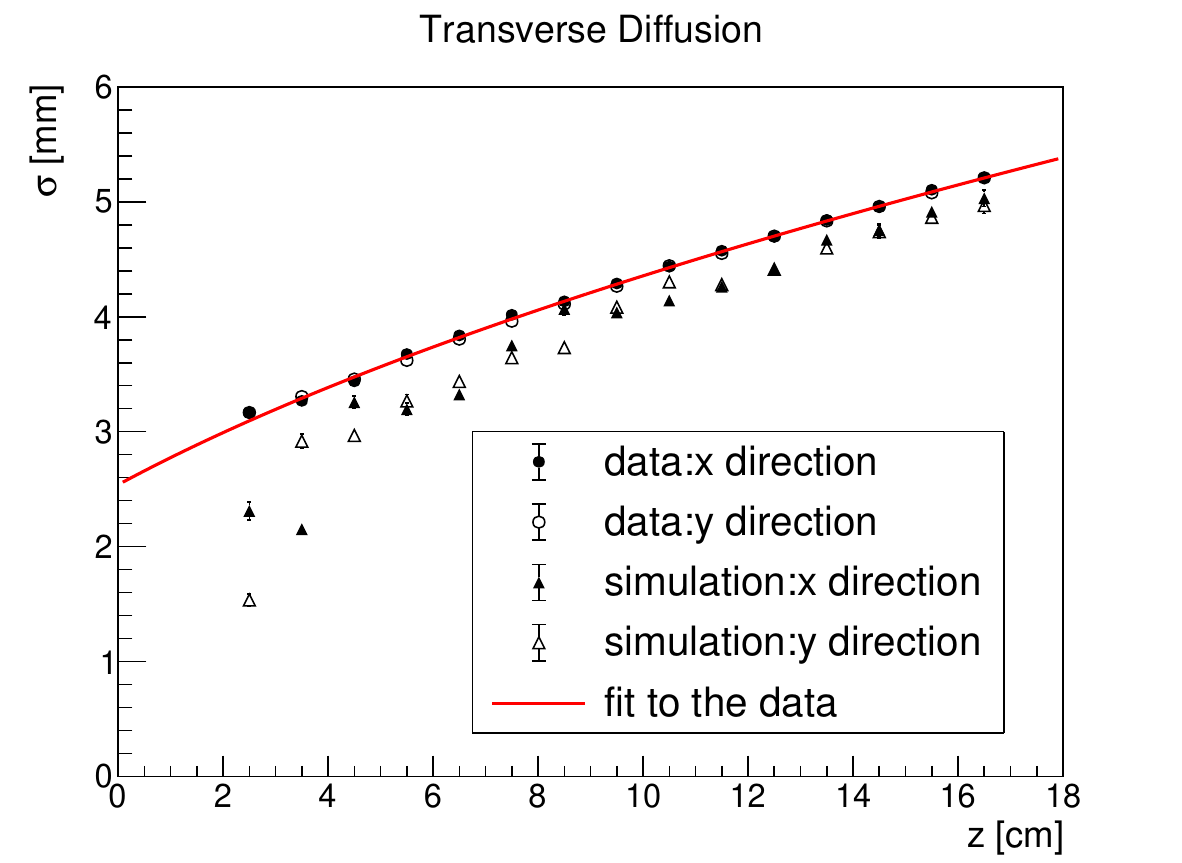}
        \end{minipage}
        \begin{minipage}{0.5\linewidth}
          \centering
          \includegraphics[width=\linewidth]{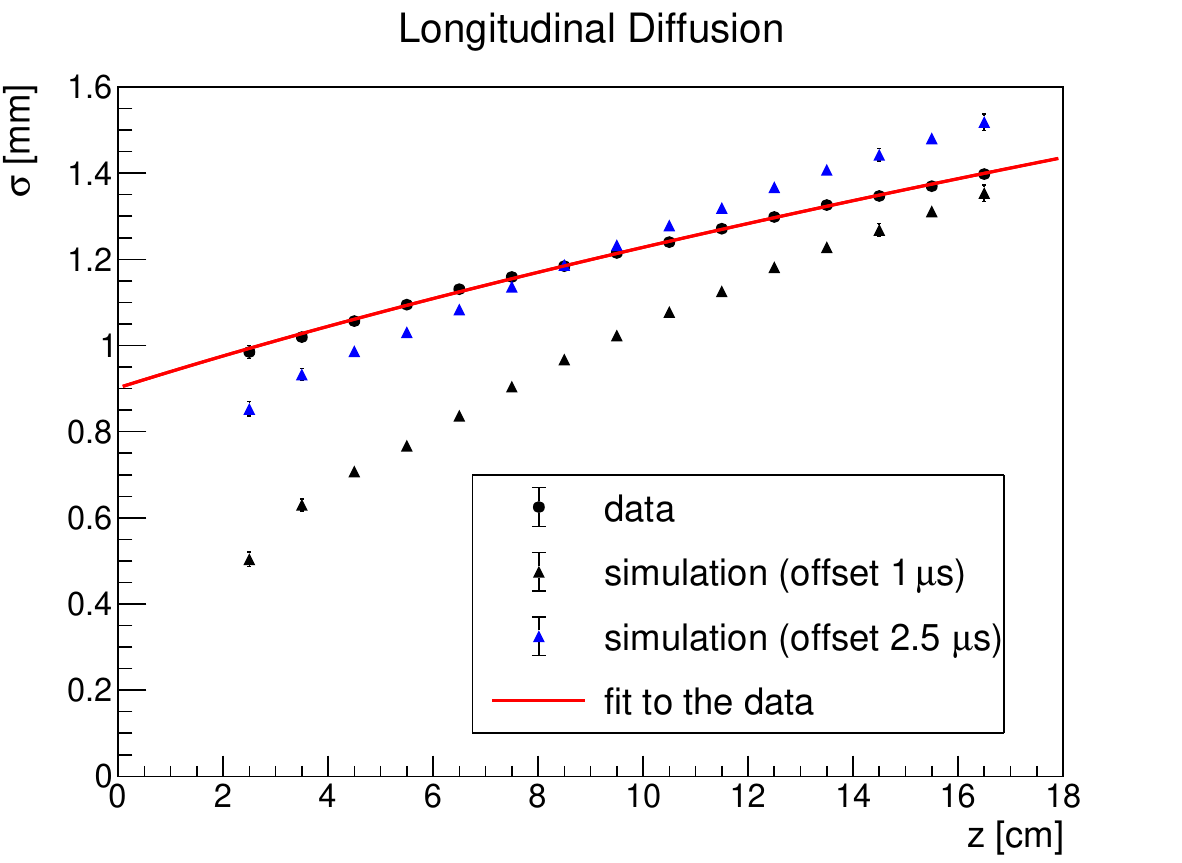}
        \end{minipage}          
        \caption{The standard deviation of the averaged hit distribution of $\mathrm{K_\alpha}$ clusters in the transverse (left panel) and the longitudinal (right panel) direction to the drift direction. The results for the simulation dataset are also shown.}
        \label{fig:diffusion}
      \end{figure}

  In the development of the algorithm to distinguish signals from backgrounds based on track images, the properties of track images should be understood and reproduced in simulation dataset.
  For this purpose, we evaluated the diffusion of the tracks.
  The $\mathrm{K_\alpha}$ clusters, whose track length is about \SI{0.8}{\mm} and much smaller than the spread by diffusion, are selected for every \SI{1}{\cm} interval in the $z$-direction and overlaid with respect to each center position to obtain averaged hit distributions.
  The standard deviations of the distribution in the $x$, $y$, and $z$ directions are plotted as a function of the $z$-position in Fig.~\ref{fig:diffusion}.
  They are fitted by the form of $\sqrt{{p_0}^2z+{p_1}^2}$, where the fit parameter $p_0$ corresponds to the diffusion of ionization electrons during drift and the parameter $p_1$ corresponds to the offset term, for example, the pixelization at ELCC for the transverse direction, and the low-pass filter in AxFEB and finite time of EL photon generation in ELCC cells for the longitudinal direction.
  The fit results are $p_0$ = \SI{0.1120+-0.0004}{\cm/\sqrt{\cm}} for the transverse direction and $p_0$ = \SI{0.0264+-0.0002}{\cm/\sqrt{\cm}} for the longitudinal direction.
  The expectations calculated by Magboltz\cite{MAGBOLTZ} are \SI{0.115}{\cm/\sqrt{\cm}} for the transverse diffusion and \SI{0.0323}{\cm/\sqrt{\cm}} for the longitudinal diffusion.
  The same analysis was performed on the simulation dataset generated with these expected diffusion constants.
  The simulation takes into account the generation of photons in the ELCC (\SI{1}{\us}) and the response of AxFEB.
  As shown in Fig.~\ref{fig:diffusion}, the transverse diffusion is roughly reproduced but the longitudinal diffusion differs both for the offset and $z$ dependence.
  For the longitudinal direction, an additional offset of \SI{1.5}{\us} is added to the simulation and is also displayed in Fig.~\ref{fig:diffusion}.
  The agreement between the measurement and simulation becomes better.
  There is, however, still disagreement, indicating that the diffusion constant is different.
  The diffusion constant is sensitive to the impurities in the gas, and this may be the reason for the disagreement.
  Simulation can be tuned using these data, which is quite important to validate the algorithms separating the $0\nu\beta\beta$ signal from the gamma-ray background based on the track image.

\section{Conclusion}
  We have upgraded the AXEL prototype detector with a \SI{180}{\L} pressure vessel and evaluated the performance at around the Q value of the $^{136}\mathrm{Xe}$ $0\nu\beta\beta$, \SI{2458}{\keV}.
  The number of the ELCC units was increased to 12.
  The structure of ELCC was also upgraded to suppress discharges.
  Data were taken at \SI{7.6}{bar} with irradiation of \SI{1836}{\keV} gamma-rays from an $^{88}\mathrm{Y}$ source and with thorium series gamma-rays including \SI{2615}{\keV} from $^{208}\mathrm{Tl}$.
  The obtained FWHM energy resolution is \SI{0.73+-0.11}{\%} at \SI{1836}{\keV}.
  The FWHM energy resolution at the Q value is estimated to be \SI{0.60+-0.03}{\%} when extrapolated by $a\sqrt{E}$, and \SI{0.70+-0.21}{\%} when extrapolated by $a\sqrt{E+bE^2}$.
  This result proves the scalability of the AXEL detector with the ELCC while maintaining high energy resolution.
  The factors that determine the energy resolution were evaluated, and it was shown that further development of the ELCC will improve the energy resolution.
  In the reconstructed track images, the blob structures are confirmed, which correspond to the number of electrons in the event.
  The diffusion constants are derived from the data, which is important to develop algorithms to discriminate the signal and background based on simulated track images.
  
\section*{Acknowledgements}
This work was supported by the JSPS KAKENHI Grant Numbers
18H05540, 
18J00365, 
18J20453, 
19K14738, 
20H00159, 
20H05251. 
We also appreciate the support for our project by Institute for Cosmic Ray Research, the University of Tokyo.
The development of the front-end board AxFEB is supported by Open-It (Open Source Consortium of Instrumentation).
\bibliographystyle{myptephy}
\bibliography{bibfile}

\end{document}